\documentclass[pra,reprint,superscriptaddress,longbibliography]{revtex4-2}

\usepackage{natbib}
\usepackage[section]{placeins}
\usepackage{pgfplots,amsfonts}
\usepackage[breaklinks]{hyperref}
\usepackage{amssymb}
\usepackage{amsmath,amsthm,bm}
\usepackage{amsfonts}
\usepackage{cleveref}
\usepackage{mathtools}
\usepackage{tikz}
\hypersetup{citecolor=red,colorlinks=true,urlcolor=blue}
\usepackage{algorithm,setspace}
\usepackage{algpseudocode}
\usetikzlibrary{angles, quotes}
\usetikzlibrary{positioning}

\usepackage[normalem]{ulem}
\usepackage{color}
\usepackage{xcolor}

\definecolor{Ugreen}{HTML}{198a11}

\begin{document}
\title{Quantum Simulation of Conical Intersections}

\author{Yuchen Wang and David A. Mazziotti}

\email{damazz@uchicago.edu}

\affiliation{Department of Chemistry and The James Franck Institute, The University of Chicago, Chicago, Illinois 60637 USA}

\date{Submitted January 27, 2024}

\begin{abstract}
We explore the simulation of conical intersections (CIs) on quantum devices, setting the groundwork for potential applications in nonadiabatic quantum dynamics within molecular systems.  The intersecting potential energy surfaces of H$_{3}^{+}$ are computed from a variance-based contracted quantum eigensolver.  We show how the CIs can be correctly described on quantum devices using wavefunctions generated by the anti-Hermitian contracted Schr{\"o}dinger equation ansatz, which is a unitary transformation of wavefunctions that preserves the topography of CIs. A hybrid quantum-classical procedure is used to locate the seam of CIs. Additionally, we discuss the quantum implementation of the adiabatic to diabatic transformation and its relation to the geometric phase effect. Results on noisy intermediate-scale quantum devices showcase the potential of quantum computers in dealing with problems in nonadiabatic chemistry.

\end{abstract}

\maketitle


\section{Introduction}
Nonadiabatic processes involve nuclear motion on multiple potential energy surfaces (PESs). These processes are ubiquitous in nature and have been studied extensively in diverse areas such as spectroscopy, solar energy conversion, chemiluminescence, photosynthesis, and photostability of biomolecules.\cite{guo2016, schwartz1996, levine2007, demekhin2013, subotnik2016, li2022, curchod2018, matsika2004, matsika2011, nguyen2016, schuurman2018, guan2021, wang2022, zhou2024} Different potential energy surfaces can intersect at regions that exhibit a conical-shaped topography, known as conical intersections (CIs).\cite{kouppel1984,yarkony1996,domcke2004,domcke2012} In the vicinity of CIs, the Born-Oppenheimer approximation which assumes adiabaticity breaks down. Systems with nonadiabaticity can undergo sudden changes in their dominant configurations at CIs, leading to the classical and well-known ``hop'' picture between different electronic states.\cite{tully1990} CIs act as highly efficient channels for converting external excitation energy, usually carried by a photon, to internal electronic energy. Their characterization is crucial to understand rich photochemistry and photobiology processes involving energy conversion.

CIs are in general difficult to treat with quantum mechanical methods for several reasons. First, from the perspective of electronic structure theory, the excited states are harder to compute than the ground state as they correspond to first-order critical points rather than the global minimum. Moreover, the nonunitary ansatz for the wavefunction employed in some methods, such as standard coupled cluster (CC) methods, gives an incorrect topography of CIs.\cite{kohn2007,gozem2014,faraji2018,thomas2021} Second, since most electronic structure programs work under the Born-Oppenheimer approximation, results obtained from these programs are not readily applicable for subsequent chemical dynamics studies, especially near CIs. The process of converting the original adiabatic electronic structure data to a diabatic representation, referred as diabatization, is an active yet non-unified field due to the non-uniqueness of quasi-diabatic representations.\cite{mead1982, ruedenberg1993, nakamura2001, eisfeld2005, opalka2013, lenzen2017, cave1996, dobbyn1997, evenhuis2004, yarkony2019, wang2019, han2020, li2023, wang2023.2, vandaele2024} Third, the dynamics of nonadiabatic systems typically require a more complex treatment than the dynamics on a single potential energy surface. For example, we need to expand wavefunctions in the basis of every diabatic state to account for effective state transition in quantum dynamics.\cite{guo2016}

Quantum computers could be a natural solution for nonadiabatic chemistry.\cite{ollitrault2020, ollitrault2021, whitlow2023quantum, Valahu.2023, Wang.2023hrm} To address some of the concerns in the last paragraph, we observe first that the gate operations are unitary, which makes it convenient to implement a unitary ansatz of wavefunctions (e.g., the unitary coupled cluster (UCC) ansatz \cite{anand2022,lee2018} or the anti-Hermitian contracted Schr{\"o}dinger equation (ACSE) ansatz\cite{mazziotti2006, mazziotti2007, mazziotti2007.2, smart2021, smart2022, Boyn.2021, Boyn.2021u94, Smart.2022w8u, Wang.2023i8g}). They offer robust and accurate solutions to the electronic structure data near CIs. In fact, for the ACSE ansatz used in this paper, classical calculations of CIs are well established.\cite{Snyder.2010, Snyder.2011u3, Snyder.2011, Snyder.2011b} Second, quantum computers are ideal tools to perform unitary and even nonunitary propagation\cite{schlimgen2021,Hu.2020} with a possible polynomial scaling advantage over classical computers where the coupling potential term can be expressed as an entanglement of encoded qubits.\cite{ollitrault2020} Third, the transformation from adiabatic wavefunctions to diabatic wavefunctions is unitary and can be easily implemented as parametric gates during state preparation on quantum computers. The geometric phase,\cite{longuet1958,Valahu.2023,Wang.2023hrm} a global phase factor dressing the wavefunctions near CIs, can also be encoded with simple rotations in the Pauli basis, which is a natural advantage of quantum computers.

In this paper we evaluate the performance of quantum computers in describing CIs. Some key issues associated with CIs, such as seam curvature, optimization and geometric phase are discussed. We implement the electronic structure simulation of $\mathrm{H_{3}^{+}}$ with and without noise using the excited-state contracted quantum eigensolver (CQE) proposed in Ref.~\cite{wang2023} in which the wavefunctions are generated by the ACSE ansatz. The theory and methodology of CI including an overview of CQE are presented in section II. Results and outlooks are further discussed in section III and IV, respectively.

\section{Theory}

\subsection{Diabatic Hamiltonian matrix and CIs}

The electronic Schr{\"o}dinger equation can be written as
\begin{equation}
 [ \mathbf{{H}}^{d}(\mathbf{R}) - \mathbf{I}E^{J}(\mathbf{R}) ]\mathbf{d}^{J}(\mathbf{R}) = 0
\end{equation}
in which $\mathbf{{H}}^{d}$ is the quasi-diabatic Hamiltonian matrix. $E^{J}$ and $\mathbf{d}^{J}$ are the adiabatic energies and wavefunctions of the J\textit{th} state respectively, corresponding to nuclear coordinate $\mathbf{R}$. We consider a two-state example in which $\mathbf{{H}}^{d}$ is a two by two matrix throughout this work, where most conclusions can be readily extended to additional electronic states. To obtain degenerate eigenvalues, we require
\begin{equation}
    (H_{11}-H_{22})^{2}+4H_{12}H_{21}=0
\end{equation}
where $H_{IJ}$ are matrix elements of $\mathbf{{H}}^{d}$.

When wavefunctions are non-Hermitian as generated from nonunitary exponential ansatz (for example, standard coupled cluster ansatz), the above equation is the only constraint to form CIs. However, this creates a nonphysical $(N-1)$ artifact that accompanies complex eigenvalues in the vicinity of true CIs, where $N$ is the molecular degree of freedom.\cite{kohn2007, gozem2014, faraji2018, thomas2021} The true CI is a submanifold of the potential energy surfaces where the following two constraint equations, one for diagonal and one for off-diagonal term,
\begin{equation}
H_{11}=H_{22}, \; H_{12}=H_{21}=0
\end{equation}
are simultaneously satisfied in the diabatic representation.

It is then easily recognized that the dimension of CIs is $(N-2)$. While the diagonal condition is easy to constrain, the off-diagonal condition is subtle because it is not directly available in electronic structure programs. For some molecules, as we will show in this paper, it is possible to find two states with high symmetry such that the couplings between them are strictly zero by symmetry, a situation known as symmetry-required CIs. The symmetry, however, does not serve as a necessary condition for the existence of CIs as some CIs occur in the more general category of ``accidental'' CIs.\cite{yarkony1996}

\subsection{Geometric phase effect}

Most modern quantum chemistry programs assume the Born-Oppenheimer approximation and thus, produce electronic structure data in the adiabatic representation.  The adiabatic representation, however, fails to describe nonadiabatic dynamics because state couplings are always zero. The diabatic representation, on the contrary, produces well-described state couplings and smooth diabatic wavefunctions. There has been significant research effort directed towards determining the transformation from the adiabatic representation to the diabatic representation.  One reason for the abundance of such diabatization techniques is that a strict diabatic representation does not exist for polyatomic molecules; hence, we refer to their states as "quasi-diabatic."\cite{mead1982} For two-state diabatization, the transformation is written as
\begin{equation}
    \label{eq:AtD}
    | \pmb{\Psi}^{d}(\mathbf{R}) \rangle = \mathbf{U}(\mathbf{R}) | \pmb{\Psi}^{a}(\mathbf{R}) \rangle
\end{equation}
in which $\mathbf{U}$ is a unitary matrix. Some literature expresses the unitary as a rotation matrix parameterized by angle $\theta$,
\begin{equation}
\label{eq:U}
\mathbf{U}(\mathbf{R}) =
\begin{pmatrix}
\cos{\theta(\mathbf{R})} & -\sin{\theta(\mathbf{R})} \\
\sin{\theta(\mathbf{R})} & \phantom{-}\cos{\theta(\mathbf{R})}
\end{pmatrix} .
\end{equation}
We remind the reader that the expression might naturally lead to the assumption that $\theta$ is a continuous function of $\mathbf{R}$, but this is not necessarily true in the presence of CIs due to the geometric phase effect. The geometric phase effect requires that wavefunctions that are transported around a path enclosing a CI acquire an additional phase factor.\cite{longuet1958, mead1992, xie2019} A geometry-dependent and state-dependent factor $e^{iA_{K}(\mathbf{R})}(K=I,J)$ must be included in the adiabatic wavefunction. The natural advantage of using qubits to represent this two-state diabatization is that they are both isomorphic to the $\mathrm{SU(2)}$ group. Indeed on quantum computers, the phase factor can be implemented as a simple rotation gate parametrized by $A_{K}(\mathbf{R})$. One of the authors has shown in previous work\cite{wang2021} that $A_{J}(\mathbf{R})$ can be evaluated from the integral below,
\begin{equation}
    A_{J}(\mathbf{R}) = \overset{\mathbf{R}}{\underset{\mathbf{R}_{0}}{\oint}} \langle \Psi^{(a)}_{I}(\mathbf{R'})| \hat{\nabla}| \Psi^{(a)}_{J}(\mathbf{R'}) \rangle d\mathbf{R'}
    \label{eq:loop}
\end{equation}
where the integrand is the well-known derivative coupling vector.\cite{yarkony1996} As this paper focuses on the topography of CIs, namely a more ``static'' description, a detailed analysis of the nonadiabatic quantum dynamics, the geometric phase factor and its implementation on quantum platforms is reserved for future work.

\subsection{Variance-based contracted quantum eigensolver}

The electronic structure calculation in this work is performed with a variance-based contracted quantum eigensolver (CQE) that has been proposed in a previous paper.\cite{wang2023} The algorithm is briefly reviewed here.

The variance (denoted as Var) of the system is defined as:
\begin{equation}
    {\rm Var}[\Psi_{m}[^{2} F_{m}]] = \langle \Psi_{m} | ( {\hat H} - E_{m} )^{2} | \Psi_{m} \rangle
\end{equation}
We minimize the variance with respect to the parametric two-body anti-hermitian operator $\hat F_{m}$, where the wavefunction at the $m^{\rm th}$ iteration is given by the unitary ansatz as
\begin{equation}
    \label{eq:cseansatz}
    | \Psi_{m} \rangle = {\rm e}^{\hat F_{m}} | \Psi_{m-1} \rangle
\end{equation}
where
\begin{equation}
    \label{eq:F2}
    {\hat F}_{m} = \sum_{pqst}{ ^{2} F^{pq;st}_{m} {\hat a}^{\dagger}_{p} {\hat a}^{\dagger}_{q} {\hat a}^{}_{t} {\hat a}^{}_{s} }
\end{equation}
in which ${\hat a}^{\dagger}_{p}$ and ${\hat a}^{}_{p}$ are the creation and annihilation operators, respectively. The key equation guiding the optimization is derived by taking the gradient of the variance with respect to $\hat F_{m}$:
\begin{equation}
    \label{eq:grad}
    \frac{ \partial {\rm Var} }{ \partial \left ( ^{2} F^{st;pq}_{m} \right ) } = 2 \langle {\Psi_{m}} | ( {\hat \Gamma}^{pq}_{st} - {}^{2} D^{pq}_{st} ) ( {\hat H} - E_{m} )^{2} | {\Psi_{m}} \rangle ,
\end{equation}
in which ${\hat \Gamma}^{pq}_{st} = {\hat a}^{\dagger}_{p} {\hat a}^{\dagger}_{q} {\hat a}^{}_{t} {\hat a}^{}_{s}$ and the equation
\begin{equation}
    ^{2} D^{pq}_{st} = \langle \Psi_{m} | {\hat \Gamma}^{pq}_{st} | \Psi_{m} \rangle.
\end{equation}
defines the elements of the 2-RDM.  Through a self-consistent update of energy and $^{2} F_{m}$, we can converge the variance to a minima which corresponds to an excited or ground state. More details including an ancillary-assisted measurement of the variance has been reported in previous work.\cite{wang2023}

We provide additional comments regarding why variance-based CQE is suitable to describe the CIs. The convergence depends on the choice of the initial guess, which can be generated from single Slater determinant or a linear combination of them. It will converge to the nearest minimum of the variance without knowledge of the lower states. Here by nearest, we mean the most similar in configuration composition. This state-specific feature can be beneficial in studying the CIs. It allows us to tackle a specific state during the slow variation of molecular geometry without concern that the adiabatic states will cross. Note this also coincides with the idea of configurational-uniformity-based diabatization as first proposed by Nakamura and Truhlar.\cite{nakamura2001}

\section{Results}

We demonstrate the approach to computing the CI with the molecule H$_{3}^{+}$.  The relative positions of the three co-planar hydrogen atoms are described in polar coordinates as $(R,0)$, $(R,\pi)$ and $(\rho, \theta)$ where $R\geq0,\rho\geq0,0\leq\theta<2\pi$, allowing us to represent the molecular geometry by the set of coordinates $(R,\rho,\theta)$. Calculations are performed with the IBMQ statevector simulator and FakeLagosV2 backend. The quantum simulation result is benchmarked with full configuration interaction calculations.  All computations are performed in the minimal Slater-type orbital (STO-3G) basis set.  Here and below we denote full configuration interaction as FCI to distinguish it from the abbreviation for the conical intersection (CI).

\subsection{Electronic structure of H$_{3}^{+}$}

The H$_{3}^{+}$ molecule exhibits arguably the simplest CI. Nonetheless, despite its simplicity, the molecule is an important species in astrochemistry, providing a useful benchmark for the study of CIs. \cite{kamisaka2002, viegas2007, barragan2006, mukherjee2014, ghosh2017, yin2021, guan2022, kwon2023} We compute the first three states of H$_{3}^{+}$ with $S_{z}=0$.
A compact mapping is used to reduce the number of required qubits to three for the first and second excited states (denoted as E$_{1}$ and E$_{2}$) of H$_{3}^{+}$. The mapping is described here. We denote the configuration state function as $| ij \rangle, (1\leq i,j\leq 3)$ with the $S_{z}=+1/2$ electron occupying the \textit{i}th molecular orbital and the $S_{z}=-1/2$ electron occupying the \textit{j}th molecular orbital. The dimension of FCI matrix is 9. A further reduction is performed by eliminating $| 11 \rangle$ by observing that it has almost no coupling to the E$_{1}$ and E$_{2}$ states. Although $| 11 \rangle$ can couple to other higher states and in principle affect the diagonalization result, the truncation has negligible effect on the energy of E$_{1}$ and E$_{2}$ ($<10^{-8}$~hartree), resulting in a total qubit number of $\log_{2} 8=3$.

\begin{figure}[t!]
    \centering
    \includegraphics[width=\linewidth]{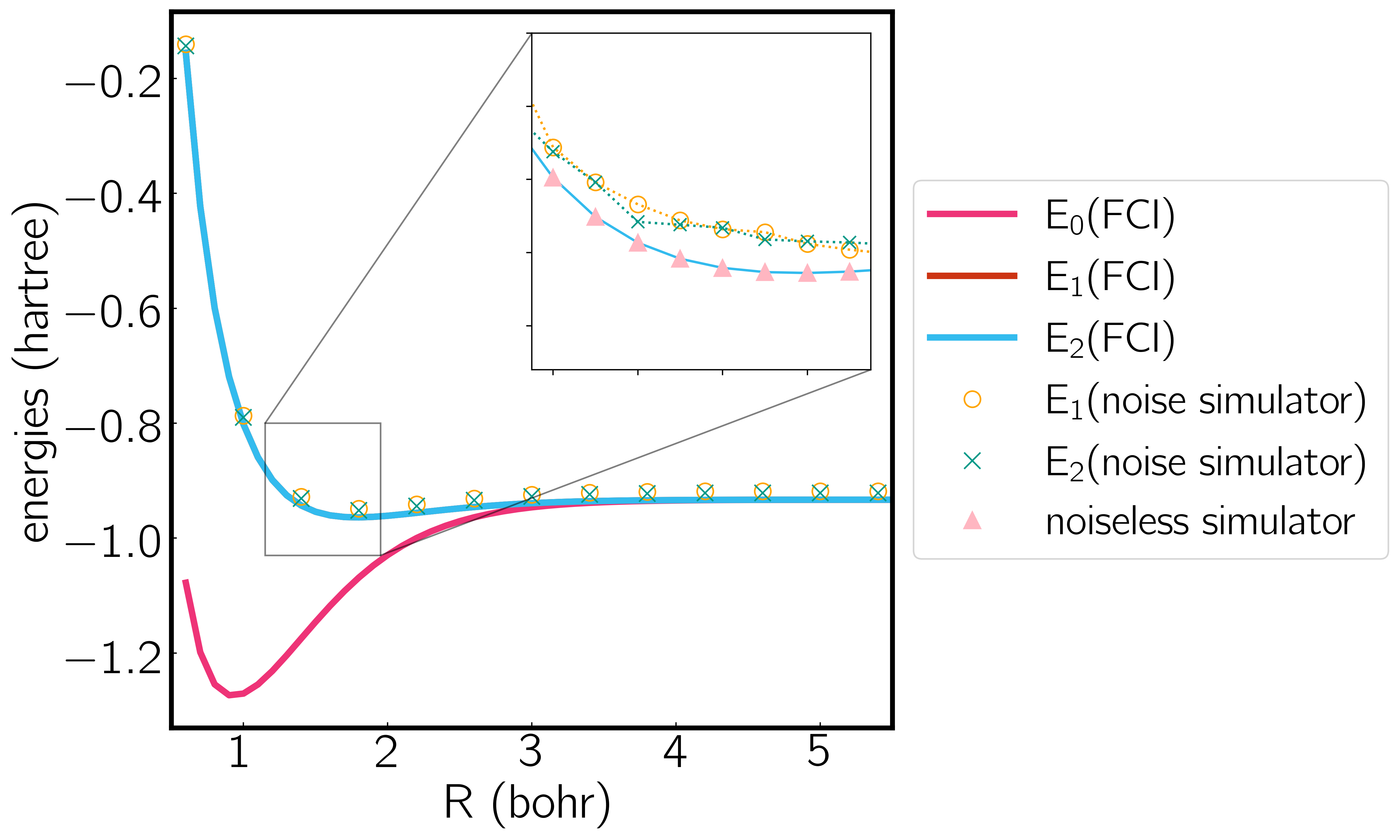}
    \caption{Potential energy curve calculated by FCI as well as noiseless and noise simulators. Molecule is treated in the D$_{3h}$ symmetry, where the polar coordinates of the three hydrogen atoms are $(R,0)$,$(R,\pi)$ and $(\sqrt{3}R, \pi/2)$. Note that E$_{1}$ and E$_{2}$ are degenerate in FCI result due to symmetry.}
    \label{fig:1}
\end{figure}

We analyze the electronic structure property of H$_{3}^{+}$ using the highly-symmetric D$_{3h}$ point group. The first and second excited states correspond to the two components of an E$^{'}$ irreducible representation and thus form the symmetry-required CIs. We plot the potential energy curve in Fig.~\ref{fig:1} obtained from the statevector simulator and a fake-backend simulator. A zoomed region of the degeneracy is given in the figure as well. The two excited states always overlap in the FCI scheme, which is consistent with our electronic structure knowledge of the system. On a noiseless statevector simulator we achieve an energy accuracy of 10$^{-6}$~hartree, where the only error comes from the trotterization, proving the exactness of the ACSE ansatz. After introducing device noise, the error for each individual state is around 12~mhartree. It is worth noting that since the error is quite uniform for both states, the error of their energy gap is significantly smaller, which is quite promising for predicting the energetic degeneracy.

\begin{figure}[t!]
    \centering
    \includegraphics[width=\linewidth]{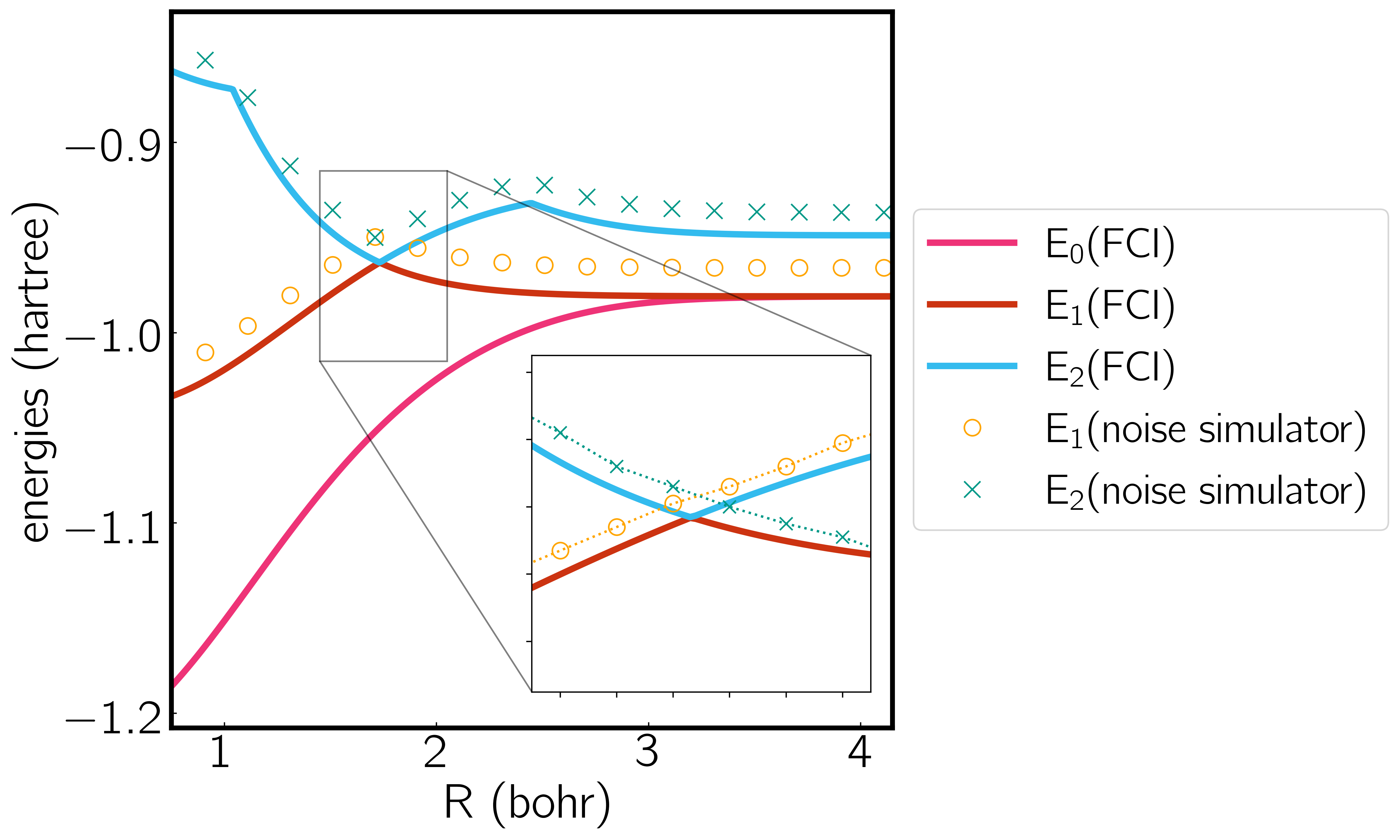}
    \caption{Potential energy curve calculated by FCI and the noise simulator. Molecule is treated in the C$_{2v}$ symmetry, where the polar coordinates of the three hydrogen atoms are $(1,0)$, $(1,\pi)$ and $(R, \pi/2)$.}
    \label{fig:2}
\end{figure}

\begin{figure}[t!]
    \centering
    \includegraphics[scale=0.3]{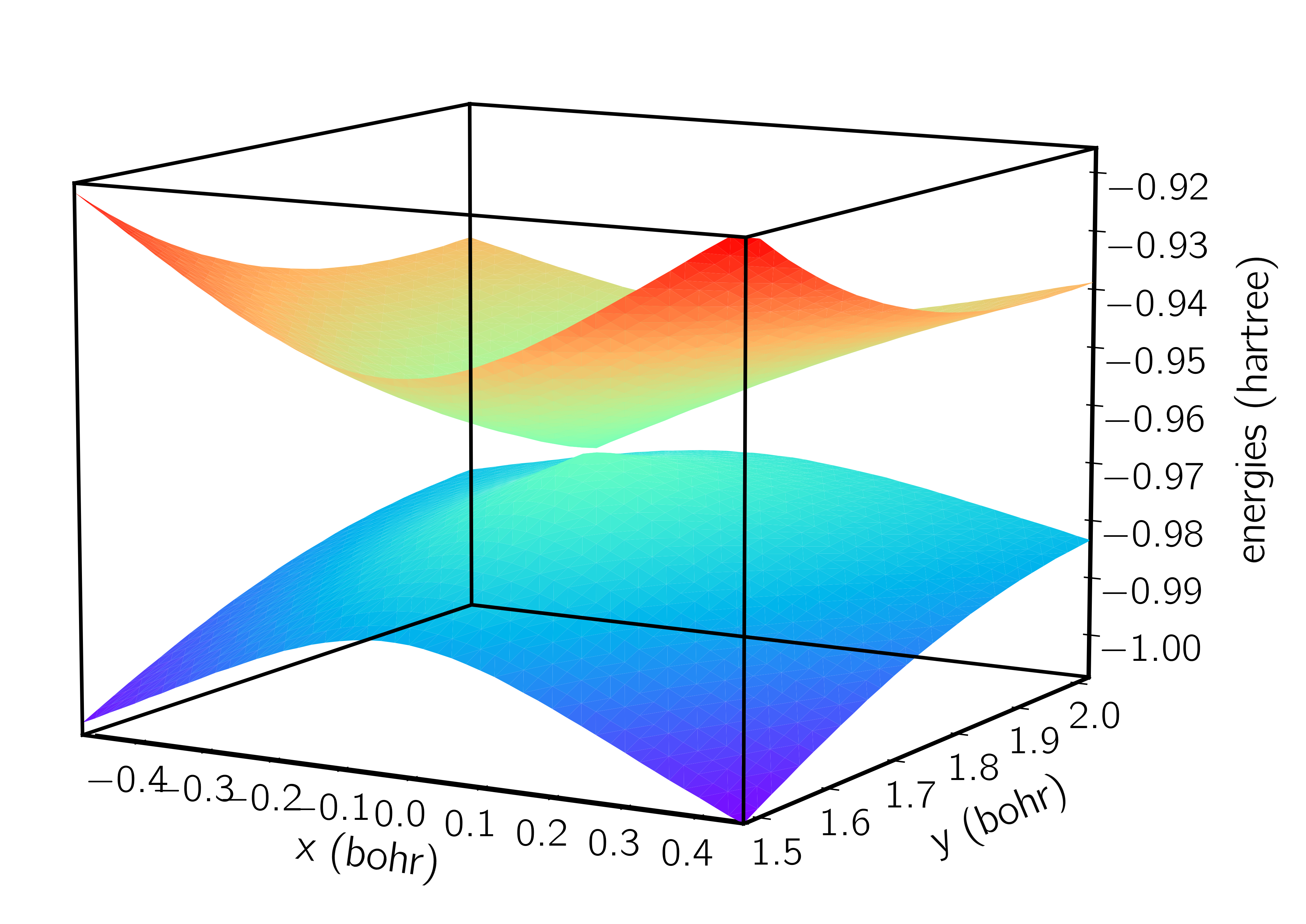}
    \caption{Intersecting adiabatic potential energy surfaces calculated from variance CQE. The grid size is 20$\times$12. Additional points are placed in the vicinity of the CI. For better illustration, we use Cartesian coordinate with the coordinates of the three atoms being (-1,0,0), (1,0,0) and (x,y,0).}
    \label{fig:3d}
\end{figure}

We next plot the dissociation curve at a lower symmetry, namely C$_{2v}$  in Fig.~\ref{fig:2}. The discontinuity of the E$_{2}$ curve is due to crossings with intruder states. We are particularly interested in the CIs between E$_{1}$ and E$_{2}$, where the two states coincide at a D$_{3h}$ geometry. Despite the relatively large error of individual states, the prediction of the location of the CIs is surprisingly accurate ($<$0.01~bohr). As mentioned before, if the error induced by noise is nearly uniform for both states and for all geometries, then the effect of noise is only to shift both potential energy surfaces by a similar amount, which should not significantly affect the topography of the CIs. To verify this, in Fig.~\ref{fig:3d} we plot the coupled potential energy surfaces as a function of the coordinates of the third hydrogen atom. It can be seen that the topography of the CIs is well reproduced. The expected cusps induced by random noise are barely discernible due to the uniformity of noise.  We note, however, that although the potential energy surfaces and relative energy gap are well reproduced, the absolute error still remains challenging on noise intermediate-scale quantum (NISQ) devices and further error mitigation techniques are needed.

\subsection{Locating the seam of CI}

The search of the minimum energy CI (MEX) is done by minimizing the constrained Lagrangian
\begin{equation}
L_{IJ}(\mathbf{R})=E_{I}(\mathbf{R})+\lambda_{0}\Delta E_{IJ}(\mathbf{R})+\sum_{k=1}^{M} \lambda_{k}C_{k}(\mathbf{R})
\end{equation}
where $C_{k}$ are certain geometry constraint equations and $\Delta E_{IJ}=0$ constrains the CIs in the adiabatic representation. In previous work, we showed that the gradient of Lagrangian corresponding to the geometry parameter set $\mathbf{R}$ can be obtained with classical calculations.\cite{wang2018} Here we use a hybrid method, where single-point calculations are performed on quantum computers and the gradient is obtained numerically and classically by varying geometries.

The dimension of CI for the triatomic molecule is only one. By varying one molecular coordinate and fixing the rest, we should obtain a one-dimensional curve that corresponds to the seam of the CI. For the special case between the E$_{1}$ and E$_{2}$ of H$_{3}^{+}$, we know that such curve is unique and corresponds to the D$_{3h}$ geometries in Fig.~1.  We report the optimization results by setting $R$ to 2.0~bohr and optimizing over the position of the third hydrogen $(r,\theta)$. To keep things simple, we used a gradient-like Newton-Raphson method with fixed step size of~0.1 where the Hessian of the Lagrangian is approximated by the identity matrix.

\begin{table}[h]
\small
    \centering
    \caption{Energy gap during a geometry optimization with respect to $\theta$ and $\rho$ starting from random guesses for these variables. The remaining parameter $R$ is fixed at 2.0~bohr.}
    \begin{tabular}{cccc}
        \hline
        Iteration Number & $\theta(^{\circ})$ & $\rho$(bohr) & $\Delta E$(hartree) \\
        \hline
        0 & 57.819 & 2.897 & 0.044\\
        1 & 58.293 & 2.364 & 0.066 \\
        2 & 71.721 & 1.741 & 0.063 \\
        3 & 81.977 & 1.741 & 0.032 \\
        4 & 92.807 & 1.756 & 0.014\\
        5 & 88.136 & 1.783 & 0.009\\
        6 & 88.537 & 1.724 & 0.005\\
        \hline
        exact & 90 & 1.732 & 0 \\
        \hline
    \end{tabular}
    \label{table1}
\end{table}

A typical optimization process is shown in Table~\ref{table1}. The performance is quite robust despite the simple setup, suggesting the gradient from quantum devices is resilient enough for geometry optimization purposes. The energy difference decreases during the optimization, except in the first iteration. This exception can occur because the Lagrangian includes contributions beyond the energy difference. An important observation is that, the noise on NISQ simulators introduces a small oscillation around our targeted D$_{3h}$ CI. The errors in the bond distance and bond angle, however, are quite small, which helps to demonstrate the accuracy of our description of the CIs.

\section{Conclusions and outlook}

Current state-of-art nonadiabatic quantum dynamics are limited to small molecules due to their exponential scaling with respect to the active vibrational modes. Quantum computers may potentially provide a solution.  In the future quantum devices with hundreds of qubits may be able to perform nonadiabatic quantum dynamics simulations that are either too expensive or intractable on high-performance classical computers. This paper provides a foundation for simulating the CIs on quantum computers, paving the way for advancements in quantum-based simulations for nonadiabatic molecular systems. Using a variance-based CQE, the energies of intersecting potential energy surfaces of molecular H$_{3}^{+}$ are accurately computed. The study achieves a correct representation of CI topography through a Hermitian wavefunction generated by the ACSE ansatz. Future work includes realizing diabatization and quantum dynamics of complex nonadiabatic molecular system on quantum devices.

\section*{Acknowledgements}
Y.W. acknowledges helpful discussion with Dr. Yafu Guan. D.A.M. gratefully acknowledges the Department of Energy, Office of Basic Energy Sciences Grant DE-SC0019215 and the U.S. National Science Foundation Grant CHE-2155082.  We acknowledge the use of IBM Quantum services for this work. The views expressed are those of the authors and do not reflect the official policy or position of IBM or the IBM Quantum team.




\bibliography{rsc} 

\begin{thebibliography}{74}%
\makeatletter
\providecommand \@ifxundefined [1]{%
 \@ifx{#1\undefined}
}%
\providecommand \@ifnum [1]{%
 \ifnum #1\expandafter \@firstoftwo
 \else \expandafter \@secondoftwo
 \fi
}%
\providecommand \@ifx [1]{%
 \ifx #1\expandafter \@firstoftwo
 \else \expandafter \@secondoftwo
 \fi
}%
\providecommand \natexlab [1]{#1}%
\providecommand \enquote  [1]{``#1''}%
\providecommand \bibnamefont  [1]{#1}%
\providecommand \bibfnamefont [1]{#1}%
\providecommand \citenamefont [1]{#1}%
\providecommand \href@noop [0]{\@secondoftwo}%
\providecommand \href [0]{\begingroup \@sanitize@url \@href}%
\providecommand \@href[1]{\@@startlink{#1}\@@href}%
\providecommand \@@href[1]{\endgroup#1\@@endlink}%
\providecommand \@sanitize@url [0]{\catcode `\\12\catcode `\$12\catcode
  `\&12\catcode `\#12\catcode `\^12\catcode `\_12\catcode `\%12\relax}%
\providecommand \@@startlink[1]{}%
\providecommand \@@endlink[0]{}%
\providecommand \url  [0]{\begingroup\@sanitize@url \@url }%
\providecommand \@url [1]{\endgroup\@href {#1}{\urlprefix }}%
\providecommand \urlprefix  [0]{URL }%
\providecommand \Eprint [0]{\href }%
\providecommand \doibase [0]{https://doi.org/}%
\providecommand \selectlanguage [0]{\@gobble}%
\providecommand \bibinfo  [0]{\@secondoftwo}%
\providecommand \bibfield  [0]{\@secondoftwo}%
\providecommand \translation [1]{[#1]}%
\providecommand \BibitemOpen [0]{}%
\providecommand \bibitemStop [0]{}%
\providecommand \bibitemNoStop [0]{.\EOS\space}%
\providecommand \EOS [0]{\spacefactor3000\relax}%
\providecommand \BibitemShut  [1]{\csname bibitem#1\endcsname}%
\let\auto@bib@innerbib\@empty
\bibitem [{\citenamefont {Guo}\ and\ \citenamefont {Yarkony}(2016)}]{guo2016}%
  \BibitemOpen
  \bibfield  {author} {\bibinfo {author} {\bibfnamefont {H.}~\bibnamefont
  {Guo}}\ and\ \bibinfo {author} {\bibfnamefont {D.~R.}\ \bibnamefont
  {Yarkony}},\ }\bibfield  {title} {\bibinfo {title} {Accurate nonadiabatic
  dynamics},\ }\href {https://doi.org/10.1039/C6CP05553B} {\bibfield  {journal}
  {\bibinfo  {journal} {Phys. Chem. Chem. Phys.}\ }\textbf {\bibinfo {volume}
  {18}},\ \bibinfo {pages} {26335} (\bibinfo {year} {2016})}\BibitemShut
  {NoStop}%
\bibitem [{\citenamefont {Schwartz}\ \emph {et~al.}(1996)\citenamefont
  {Schwartz}, \citenamefont {Bittner}, \citenamefont {Prezhdo},\ and\
  \citenamefont {Rossky}}]{schwartz1996}%
  \BibitemOpen
  \bibfield  {author} {\bibinfo {author} {\bibfnamefont {B.~J.}\ \bibnamefont
  {Schwartz}}, \bibinfo {author} {\bibfnamefont {E.~R.}\ \bibnamefont
  {Bittner}}, \bibinfo {author} {\bibfnamefont {O.~V.}\ \bibnamefont
  {Prezhdo}},\ and\ \bibinfo {author} {\bibfnamefont {P.~J.}\ \bibnamefont
  {Rossky}},\ }\bibfield  {title} {\bibinfo {title} {Quantum decoherence and
  the isotope effect in condensed phase nonadiabatic molecular dynamics
  simulations},\ }\href {https://doi.org/10.1063/1.471326} {\bibfield
  {journal} {\bibinfo  {journal} {J. Chem. Phys.}\ }\textbf {\bibinfo {volume}
  {104}},\ \bibinfo {pages} {5942–5955} (\bibinfo {year} {1996})}\BibitemShut
  {NoStop}%
\bibitem [{\citenamefont {Levine}\ and\ \citenamefont
  {Mart{\'\i}nez}(2007)}]{levine2007}%
  \BibitemOpen
  \bibfield  {author} {\bibinfo {author} {\bibfnamefont {B.~G.}\ \bibnamefont
  {Levine}}\ and\ \bibinfo {author} {\bibfnamefont {T.~J.}\ \bibnamefont
  {Mart{\'\i}nez}},\ }\bibfield  {title} {\bibinfo {title} {Isomerization
  through conical intersections},\ }\href
  {https://doi.org/10.1146/annurev.physchem.57.032905.104612} {\bibfield
  {journal} {\bibinfo  {journal} {Annu. Rev. Phys. Chem.}\ }\textbf {\bibinfo
  {volume} {58}},\ \bibinfo {pages} {613} (\bibinfo {year} {2007})}\BibitemShut
  {NoStop}%
\bibitem [{\citenamefont {Demekhin}\ and\ \citenamefont
  {Cederbaum}(2013)}]{demekhin2013}%
  \BibitemOpen
  \bibfield  {author} {\bibinfo {author} {\bibfnamefont {P.~V.}\ \bibnamefont
  {Demekhin}}\ and\ \bibinfo {author} {\bibfnamefont {L.~S.}\ \bibnamefont
  {Cederbaum}},\ }\bibfield  {title} {\bibinfo {title} {Light-induced conical
  intersections in polyatomic molecules: General theory, strategies of
  exploitation, and application},\ }\href {https://doi.org/10.1063/1.4826172}
  {\bibfield  {journal} {\bibinfo  {journal} {J. Chem. Phys.}\ }\textbf
  {\bibinfo {volume} {139}},\ \bibinfo {pages} {154314} (\bibinfo {year}
  {2013})}\BibitemShut {NoStop}%
\bibitem [{\citenamefont {Subotnik}\ \emph {et~al.}(2016)\citenamefont
  {Subotnik}, \citenamefont {Jain}, \citenamefont {Landry}, \citenamefont
  {Petit}, \citenamefont {Ouyang},\ and\ \citenamefont
  {Bellonzi}}]{subotnik2016}%
  \BibitemOpen
  \bibfield  {author} {\bibinfo {author} {\bibfnamefont {J.~E.}\ \bibnamefont
  {Subotnik}}, \bibinfo {author} {\bibfnamefont {A.}~\bibnamefont {Jain}},
  \bibinfo {author} {\bibfnamefont {B.}~\bibnamefont {Landry}}, \bibinfo
  {author} {\bibfnamefont {A.}~\bibnamefont {Petit}}, \bibinfo {author}
  {\bibfnamefont {W.}~\bibnamefont {Ouyang}},\ and\ \bibinfo {author}
  {\bibfnamefont {N.}~\bibnamefont {Bellonzi}},\ }\bibfield  {title} {\bibinfo
  {title} {Understanding the surface hopping view of electronic transitions and
  decoherence},\ }\href
  {https://doi.org/10.1146/annurev-physchem-040215-112245} {\bibfield
  {journal} {\bibinfo  {journal} {Annu. Rev. Phys. Chem.}\ }\textbf {\bibinfo
  {volume} {67}},\ \bibinfo {pages} {387} (\bibinfo {year} {2016})}\BibitemShut
  {NoStop}%
\bibitem [{\citenamefont {Li}\ \emph {et~al.}(2022)\citenamefont {Li},
  \citenamefont {Cui}, \citenamefont {Subotnik},\ and\ \citenamefont
  {Nitzan}}]{li2022}%
  \BibitemOpen
  \bibfield  {author} {\bibinfo {author} {\bibfnamefont {T.~E.}\ \bibnamefont
  {Li}}, \bibinfo {author} {\bibfnamefont {B.}~\bibnamefont {Cui}}, \bibinfo
  {author} {\bibfnamefont {J.~E.}\ \bibnamefont {Subotnik}},\ and\ \bibinfo
  {author} {\bibfnamefont {A.}~\bibnamefont {Nitzan}},\ }\bibfield  {title}
  {\bibinfo {title} {Molecular polaritonics: Chemical dynamics under strong
  light--matter coupling},\ }\href
  {https://doi.org/10.1146/annurev-physchem-090519-042621} {\bibfield
  {journal} {\bibinfo  {journal} {Annu. Rev. Phys. Chem.}\ }\textbf {\bibinfo
  {volume} {73}},\ \bibinfo {pages} {43} (\bibinfo {year} {2022})}\BibitemShut
  {NoStop}%
\bibitem [{\citenamefont {Curchod}\ and\ \citenamefont
  {Mart{\'\i}nez}(2018)}]{curchod2018}%
  \BibitemOpen
  \bibfield  {author} {\bibinfo {author} {\bibfnamefont {B.~F.}\ \bibnamefont
  {Curchod}}\ and\ \bibinfo {author} {\bibfnamefont {T.~J.}\ \bibnamefont
  {Mart{\'\i}nez}},\ }\bibfield  {title} {\bibinfo {title} {Ab initio
  nonadiabatic quantum molecular dynamics},\ }\href
  {https://doi.org/10.1021/acs.chemrev.7b00423} {\bibfield  {journal} {\bibinfo
   {journal} {Chem. Rev.}\ }\textbf {\bibinfo {volume} {118}},\ \bibinfo
  {pages} {3305} (\bibinfo {year} {2018})}\BibitemShut {NoStop}%
\bibitem [{\citenamefont {Matsika}(2004)}]{matsika2004}%
  \BibitemOpen
  \bibfield  {author} {\bibinfo {author} {\bibfnamefont {S.}~\bibnamefont
  {Matsika}},\ }\bibfield  {title} {\bibinfo {title} {Radiationless decay of
  excited states of uracil through conical intersections},\ }\href
  {https://doi.org/10.1021/jp048284n} {\bibfield  {journal} {\bibinfo
  {journal} {J. Phys. Chem. A}\ }\textbf {\bibinfo {volume} {108}},\ \bibinfo
  {pages} {7584} (\bibinfo {year} {2004})}\BibitemShut {NoStop}%
\bibitem [{\citenamefont {Matsika}\ and\ \citenamefont
  {Krause}(2011)}]{matsika2011}%
  \BibitemOpen
  \bibfield  {author} {\bibinfo {author} {\bibfnamefont {S.}~\bibnamefont
  {Matsika}}\ and\ \bibinfo {author} {\bibfnamefont {P.}~\bibnamefont
  {Krause}},\ }\bibfield  {title} {\bibinfo {title} {Nonadiabatic events and
  conical intersections},\ }\href
  {https://doi.org/10.1146/annurev-physchem-032210-103450} {\bibfield
  {journal} {\bibinfo  {journal} {Annu. Rev. Phys. Chem.}\ }\textbf {\bibinfo
  {volume} {62}},\ \bibinfo {pages} {621} (\bibinfo {year} {2011})}\BibitemShut
  {NoStop}%
\bibitem [{\citenamefont {Nguyen}\ \emph {et~al.}(2016)\citenamefont {Nguyen},
  \citenamefont {Spata},\ and\ \citenamefont {Matsika}}]{nguyen2016}%
  \BibitemOpen
  \bibfield  {author} {\bibinfo {author} {\bibfnamefont {Q.~L.}\ \bibnamefont
  {Nguyen}}, \bibinfo {author} {\bibfnamefont {V.~A.}\ \bibnamefont {Spata}},\
  and\ \bibinfo {author} {\bibfnamefont {S.}~\bibnamefont {Matsika}},\
  }\bibfield  {title} {\bibinfo {title} {Photophysical properties of
  pyrrolocytosine, a cytosine fluorescent base analogue},\ }\href
  {https://doi.org/10.1039/C6CP01559J} {\bibfield  {journal} {\bibinfo
  {journal} {Phys. Chem. Chem. Phys.}\ }\textbf {\bibinfo {volume} {18}},\
  \bibinfo {pages} {20189} (\bibinfo {year} {2016})}\BibitemShut {NoStop}%
\bibitem [{\citenamefont {Schuurman}\ and\ \citenamefont
  {Stolow}(2018)}]{schuurman2018}%
  \BibitemOpen
  \bibfield  {author} {\bibinfo {author} {\bibfnamefont {M.~S.}\ \bibnamefont
  {Schuurman}}\ and\ \bibinfo {author} {\bibfnamefont {A.}~\bibnamefont
  {Stolow}},\ }\bibfield  {title} {\bibinfo {title} {Dynamics at conical
  intersections},\ }\href
  {https://doi.org/10.1146/annurev-physchem-052516-050721} {\bibfield
  {journal} {\bibinfo  {journal} {Annu. Rev. Phys. Chem.}\ }\textbf {\bibinfo
  {volume} {69}},\ \bibinfo {pages} {427} (\bibinfo {year} {2018})}\BibitemShut
  {NoStop}%
\bibitem [{\citenamefont {Guan}\ \emph {et~al.}(2021)\citenamefont {Guan},
  \citenamefont {Xie}, \citenamefont {Yarkony},\ and\ \citenamefont
  {Guo}}]{guan2021}%
  \BibitemOpen
  \bibfield  {author} {\bibinfo {author} {\bibfnamefont {Y.}~\bibnamefont
  {Guan}}, \bibinfo {author} {\bibfnamefont {C.}~\bibnamefont {Xie}}, \bibinfo
  {author} {\bibfnamefont {D.~R.}\ \bibnamefont {Yarkony}},\ and\ \bibinfo
  {author} {\bibfnamefont {H.}~\bibnamefont {Guo}},\ }\bibfield  {title}
  {\bibinfo {title} {High-fidelity first principles nonadiabaticity:
  diabatization, analytic representation of global diabatic potential energy
  matrices, and quantum dynamics},\ }\href {https://doi.org/10.1039/D1CP03008F}
  {\bibfield  {journal} {\bibinfo  {journal} {Phys. Chem. Chem. Phys.}\
  }\textbf {\bibinfo {volume} {23}},\ \bibinfo {pages} {24962} (\bibinfo {year}
  {2021})}\BibitemShut {NoStop}%
\bibitem [{\citenamefont {Wang}\ \emph {et~al.}(2022)\citenamefont {Wang},
  \citenamefont {Guo},\ and\ \citenamefont {Yarkony}}]{wang2022}%
  \BibitemOpen
  \bibfield  {author} {\bibinfo {author} {\bibfnamefont {Y.}~\bibnamefont
  {Wang}}, \bibinfo {author} {\bibfnamefont {H.}~\bibnamefont {Guo}},\ and\
  \bibinfo {author} {\bibfnamefont {D.~R.}\ \bibnamefont {Yarkony}},\
  }\bibfield  {title} {\bibinfo {title} {Internal conversion and intersystem
  crossing dynamics based on coupled potential energy surfaces with full
  geometry-dependent spin--orbit and derivative couplings. nonadiabatic
  photodissociation dynamics of {NH}$_{3}$({A}) leading to the
  {NH}({X$^{3}\Sigma^{-}$},a$^{1}${$\Delta$})+{H}$_{2}$ channel},\ }\href
  {https://doi.org/10.1039/D2CP01271E} {\bibfield  {journal} {\bibinfo
  {journal} {Phys. Chem. Chem. Phys.}\ }\textbf {\bibinfo {volume} {24}},\
  \bibinfo {pages} {15060} (\bibinfo {year} {2022})}\BibitemShut {NoStop}%
\bibitem [{\citenamefont {Zhou}\ \emph {et~al.}(2024)\citenamefont {Zhou},
  \citenamefont {Shu}, \citenamefont {Wang}, \citenamefont {Leszczynski},\ and\
  \citenamefont {Prezhdo}}]{zhou2024}%
  \BibitemOpen
  \bibfield  {author} {\bibinfo {author} {\bibfnamefont {J.-G.}\ \bibnamefont
  {Zhou}}, \bibinfo {author} {\bibfnamefont {Y.}~\bibnamefont {Shu}}, \bibinfo
  {author} {\bibfnamefont {Y.}~\bibnamefont {Wang}}, \bibinfo {author}
  {\bibfnamefont {J.}~\bibnamefont {Leszczynski}},\ and\ \bibinfo {author}
  {\bibfnamefont {O.}~\bibnamefont {Prezhdo}},\ }\bibfield  {title} {\bibinfo
  {title} {Dissociation time, quantum yield and dynamic reaction pathways in
  the thermolysis of {Trans-3,4-dimethyl-1,2-dioxetane}},\ }\href@noop {}
  {\bibfield  {journal} {\bibinfo  {journal} {J. Phys. Chem. Lett.}\ ,\
  \bibinfo {pages} {submitted}} (\bibinfo {year} {2024})}\BibitemShut {NoStop}%
\bibitem [{\citenamefont {K{\"o}uppel}\ \emph {et~al.}(1984)\citenamefont
  {K{\"o}uppel}, \citenamefont {Domcke},\ and\ \citenamefont
  {Cederbaum}}]{kouppel1984}%
  \BibitemOpen
  \bibfield  {author} {\bibinfo {author} {\bibfnamefont {H.}~\bibnamefont
  {K{\"o}uppel}}, \bibinfo {author} {\bibfnamefont {W.}~\bibnamefont
  {Domcke}},\ and\ \bibinfo {author} {\bibfnamefont {L.}~\bibnamefont
  {Cederbaum}},\ }\bibfield  {title} {\bibinfo {title} {Multimode molecular
  dynamics beyond the {B}orn-{O}ppenheimer approximation},\ }\href@noop {}
  {\bibfield  {journal} {\bibinfo  {journal} {Adv. Chem. Phys.}\ ,\ \bibinfo
  {pages} {59}} (\bibinfo {year} {1984})}\BibitemShut {NoStop}%
\bibitem [{\citenamefont {Yarkony}(1996)}]{yarkony1996}%
  \BibitemOpen
  \bibfield  {author} {\bibinfo {author} {\bibfnamefont {D.~R.}\ \bibnamefont
  {Yarkony}},\ }\bibfield  {title} {\bibinfo {title} {Diabolical conical
  intersections},\ }\href {https://doi.org/10.1103/RevModPhys.68.985}
  {\bibfield  {journal} {\bibinfo  {journal} {Rev. Mod. Phys.}\ }\textbf
  {\bibinfo {volume} {68}},\ \bibinfo {pages} {985} (\bibinfo {year}
  {1996})}\BibitemShut {NoStop}%
\bibitem [{\citenamefont {Domcke}\ \emph {et~al.}(2004)\citenamefont {Domcke},
  \citenamefont {Yarkony},\ and\ \citenamefont {K{\"o}ppel}}]{domcke2004}%
  \BibitemOpen
  \bibfield  {author} {\bibinfo {author} {\bibfnamefont {W.}~\bibnamefont
  {Domcke}}, \bibinfo {author} {\bibfnamefont {D.}~\bibnamefont {Yarkony}},\
  and\ \bibinfo {author} {\bibfnamefont {H.}~\bibnamefont {K{\"o}ppel}},\
  }\href@noop {} {\emph {\bibinfo {title} {Conical intersections: Electronic
  structure, dynamics \& spectroscopy}}},\ Vol.~\bibinfo {volume} {15}\
  (\bibinfo  {publisher} {World Scientific},\ \bibinfo {year}
  {2004})\BibitemShut {NoStop}%
\bibitem [{\citenamefont {Domcke}\ and\ \citenamefont
  {Yarkony}(2012)}]{domcke2012}%
  \BibitemOpen
  \bibfield  {author} {\bibinfo {author} {\bibfnamefont {W.}~\bibnamefont
  {Domcke}}\ and\ \bibinfo {author} {\bibfnamefont {D.~R.}\ \bibnamefont
  {Yarkony}},\ }\bibfield  {title} {\bibinfo {title} {Role of conical
  intersections in molecular spectroscopy and photoinduced chemical dynamics},\
  }\href {https://doi.org/10.1146/annurev-physchem-032210-103522} {\bibfield
  {journal} {\bibinfo  {journal} {Annu. Rev. Phys. Chem}\ }\textbf {\bibinfo
  {volume} {63}},\ \bibinfo {pages} {325} (\bibinfo {year} {2012})}\BibitemShut
  {NoStop}%
\bibitem [{\citenamefont {Tully}(1990)}]{tully1990}%
  \BibitemOpen
  \bibfield  {author} {\bibinfo {author} {\bibfnamefont {J.~C.}\ \bibnamefont
  {Tully}},\ }\bibfield  {title} {\bibinfo {title} {Molecular dynamics with
  electronic transitions},\ }\href {https://doi.org/10.1063/1.459170}
  {\bibfield  {journal} {\bibinfo  {journal} {J. Chem. Phys.}\ }\textbf
  {\bibinfo {volume} {93}},\ \bibinfo {pages} {1061} (\bibinfo {year}
  {1990})}\BibitemShut {NoStop}%
\bibitem [{\citenamefont {K{\"o}hn}\ and\ \citenamefont
  {Tajti}(2007)}]{kohn2007}%
  \BibitemOpen
  \bibfield  {author} {\bibinfo {author} {\bibfnamefont {A.}~\bibnamefont
  {K{\"o}hn}}\ and\ \bibinfo {author} {\bibfnamefont {A.}~\bibnamefont
  {Tajti}},\ }\bibfield  {title} {\bibinfo {title} {Can coupled-cluster theory
  treat conical intersections?},\ }\href {https://doi.org/10.1063/1.2755681}
  {\bibfield  {journal} {\bibinfo  {journal} {J. Chem. Phys.}\ }\textbf
  {\bibinfo {volume} {127}},\ \bibinfo {pages} {044105} (\bibinfo {year}
  {2007})}\BibitemShut {NoStop}%
\bibitem [{\citenamefont {Gozem}\ \emph {et~al.}(2014)\citenamefont {Gozem},
  \citenamefont {Melaccio}, \citenamefont {Valentini}, \citenamefont {Filatov},
  \citenamefont {Huix-Rotllant}, \citenamefont {Ferre}, \citenamefont {Frutos},
  \citenamefont {Angeli}, \citenamefont {Krylov}, \citenamefont {Granovsky}
  \emph {et~al.}}]{gozem2014}%
  \BibitemOpen
  \bibfield  {author} {\bibinfo {author} {\bibfnamefont {S.}~\bibnamefont
  {Gozem}}, \bibinfo {author} {\bibfnamefont {F.}~\bibnamefont {Melaccio}},
  \bibinfo {author} {\bibfnamefont {A.}~\bibnamefont {Valentini}}, \bibinfo
  {author} {\bibfnamefont {M.}~\bibnamefont {Filatov}}, \bibinfo {author}
  {\bibfnamefont {M.}~\bibnamefont {Huix-Rotllant}}, \bibinfo {author}
  {\bibfnamefont {N.}~\bibnamefont {Ferre}}, \bibinfo {author} {\bibfnamefont
  {L.~M.}\ \bibnamefont {Frutos}}, \bibinfo {author} {\bibfnamefont
  {C.}~\bibnamefont {Angeli}}, \bibinfo {author} {\bibfnamefont {A.~I.}\
  \bibnamefont {Krylov}}, \bibinfo {author} {\bibfnamefont {A.~A.}\
  \bibnamefont {Granovsky}}, \emph {et~al.},\ }\bibfield  {title} {\bibinfo
  {title} {Shape of multireference, equation-of-motion coupled-cluster, and
  density functional theory potential energy surfaces at a conical
  intersection},\ }\href {https://doi.org/10.1021/ct500154k} {\bibfield
  {journal} {\bibinfo  {journal} {J. Chem. Theory Comput.}\ }\textbf {\bibinfo
  {volume} {10}},\ \bibinfo {pages} {3074} (\bibinfo {year}
  {2014})}\BibitemShut {NoStop}%
\bibitem [{\citenamefont {Faraji}\ \emph {et~al.}(2018)\citenamefont {Faraji},
  \citenamefont {Matsika},\ and\ \citenamefont {Krylov}}]{faraji2018}%
  \BibitemOpen
  \bibfield  {author} {\bibinfo {author} {\bibfnamefont {S.}~\bibnamefont
  {Faraji}}, \bibinfo {author} {\bibfnamefont {S.}~\bibnamefont {Matsika}},\
  and\ \bibinfo {author} {\bibfnamefont {A.~I.}\ \bibnamefont {Krylov}},\
  }\bibfield  {title} {\bibinfo {title} {Calculations of non-adiabatic
  couplings within equation-of-motion coupled-cluster framework: Theory,
  implementation, and validation against multi-reference methods},\ }\href
  {https://doi.org/10.1063/1.5009433} {\bibfield  {journal} {\bibinfo
  {journal} {J. Chem. Phys.}\ }\textbf {\bibinfo {volume} {148}},\ \bibinfo
  {pages} {044103} (\bibinfo {year} {2018})}\BibitemShut {NoStop}%
\bibitem [{\citenamefont {Thomas}\ \emph {et~al.}(2021)\citenamefont {Thomas},
  \citenamefont {Hampe}, \citenamefont {Stopkowicz},\ and\ \citenamefont
  {Gauss}}]{thomas2021}%
  \BibitemOpen
  \bibfield  {author} {\bibinfo {author} {\bibfnamefont {S.}~\bibnamefont
  {Thomas}}, \bibinfo {author} {\bibfnamefont {F.}~\bibnamefont {Hampe}},
  \bibinfo {author} {\bibfnamefont {S.}~\bibnamefont {Stopkowicz}},\ and\
  \bibinfo {author} {\bibfnamefont {J.}~\bibnamefont {Gauss}},\ }\bibfield
  {title} {\bibinfo {title} {Complex ground-state and excitation energies in
  coupled-cluster theory},\ }\href
  {https://doi.org/10.1080/00268976.2021.1968056} {\bibfield  {journal}
  {\bibinfo  {journal} {Mol. Phys.}\ }\textbf {\bibinfo {volume} {119}},\
  \bibinfo {pages} {e1968056} (\bibinfo {year} {2021})}\BibitemShut {NoStop}%
\bibitem [{\citenamefont {Mead}\ and\ \citenamefont
  {Truhlar}(1982)}]{mead1982}%
  \BibitemOpen
  \bibfield  {author} {\bibinfo {author} {\bibfnamefont {C.~A.}\ \bibnamefont
  {Mead}}\ and\ \bibinfo {author} {\bibfnamefont {D.~G.}\ \bibnamefont
  {Truhlar}},\ }\bibfield  {title} {\bibinfo {title} {Conditions for the
  definition of a strictly diabatic electronic basis for molecular systems},\
  }\href {https://doi.org/10.1063/1.443853} {\bibfield  {journal} {\bibinfo
  {journal} {J. Chem. Phys.}\ }\textbf {\bibinfo {volume} {77}},\ \bibinfo
  {pages} {6090} (\bibinfo {year} {1982})}\BibitemShut {NoStop}%
\bibitem [{\citenamefont {Ruedenberg}\ and\ \citenamefont
  {Atchity}(1993)}]{ruedenberg1993}%
  \BibitemOpen
  \bibfield  {author} {\bibinfo {author} {\bibfnamefont {K.}~\bibnamefont
  {Ruedenberg}}\ and\ \bibinfo {author} {\bibfnamefont {G.~J.}\ \bibnamefont
  {Atchity}},\ }\bibfield  {title} {\bibinfo {title} {A quantum chemical
  determination of diabatic states},\ }\href {https://doi.org/10.1063/1.466125}
  {\bibfield  {journal} {\bibinfo  {journal} {J. Chem. Phys.}\ }\textbf
  {\bibinfo {volume} {99}},\ \bibinfo {pages} {3799} (\bibinfo {year}
  {1993})}\BibitemShut {NoStop}%
\bibitem [{\citenamefont {Nakamura}\ and\ \citenamefont
  {Truhlar}(2001)}]{nakamura2001}%
  \BibitemOpen
  \bibfield  {author} {\bibinfo {author} {\bibfnamefont {H.}~\bibnamefont
  {Nakamura}}\ and\ \bibinfo {author} {\bibfnamefont {D.~G.}\ \bibnamefont
  {Truhlar}},\ }\bibfield  {title} {\bibinfo {title} {The direct calculation of
  diabatic states based on configurational uniformity},\ }\href
  {https://doi.org/10.1063/1.1412879} {\bibfield  {journal} {\bibinfo
  {journal} {J. Chem. Phys.}\ }\textbf {\bibinfo {volume} {115}},\ \bibinfo
  {pages} {10353} (\bibinfo {year} {2001})}\BibitemShut {NoStop}%
\bibitem [{\citenamefont {Eisfeld}\ and\ \citenamefont
  {Viel}(2005)}]{eisfeld2005}%
  \BibitemOpen
  \bibfield  {author} {\bibinfo {author} {\bibfnamefont {W.}~\bibnamefont
  {Eisfeld}}\ and\ \bibinfo {author} {\bibfnamefont {A.}~\bibnamefont {Viel}},\
  }\bibfield  {title} {\bibinfo {title} {Higher order {A+E} {$\otimes$} e
  pseudo-{J}ahn--{T}eller coupling},\ }\href
  {https://doi.org/10.1063/1.1904594} {\bibfield  {journal} {\bibinfo
  {journal} {J. Chem. Phys.}\ }\textbf {\bibinfo {volume} {122}},\ \bibinfo
  {pages} {204317} (\bibinfo {year} {2005})}\BibitemShut {NoStop}%
\bibitem [{\citenamefont {Opalka}\ and\ \citenamefont
  {Domcke}(2013)}]{opalka2013}%
  \BibitemOpen
  \bibfield  {author} {\bibinfo {author} {\bibfnamefont {D.}~\bibnamefont
  {Opalka}}\ and\ \bibinfo {author} {\bibfnamefont {W.}~\bibnamefont
  {Domcke}},\ }\bibfield  {title} {\bibinfo {title} {Interpolation of
  multi-sheeted multi-dimensional potential-energy surfaces via a linear
  optimization procedure},\ }\href {https://doi.org/10.1063/1.4808358}
  {\bibfield  {journal} {\bibinfo  {journal} {J. Chem. Phys.}\ }\textbf
  {\bibinfo {volume} {138}},\ \bibinfo {pages} {224103} (\bibinfo {year}
  {2013})}\BibitemShut {NoStop}%
\bibitem [{\citenamefont {Lenzen}\ and\ \citenamefont
  {Manthe}(2017)}]{lenzen2017}%
  \BibitemOpen
  \bibfield  {author} {\bibinfo {author} {\bibfnamefont {T.}~\bibnamefont
  {Lenzen}}\ and\ \bibinfo {author} {\bibfnamefont {U.}~\bibnamefont
  {Manthe}},\ }\bibfield  {title} {\bibinfo {title} {Neural network based
  coupled diabatic potential energy surfaces for reactive scattering},\ }\href
  {https://doi.org/10.1063/1.4997995} {\bibfield  {journal} {\bibinfo
  {journal} {J. Chem. Phys.}\ }\textbf {\bibinfo {volume} {147}},\ \bibinfo
  {pages} {084105} (\bibinfo {year} {2017})}\BibitemShut {NoStop}%
\bibitem [{\citenamefont {Cave}\ and\ \citenamefont {Newton}(1996)}]{cave1996}%
  \BibitemOpen
  \bibfield  {author} {\bibinfo {author} {\bibfnamefont {R.~J.}\ \bibnamefont
  {Cave}}\ and\ \bibinfo {author} {\bibfnamefont {M.~D.}\ \bibnamefont
  {Newton}},\ }\bibfield  {title} {\bibinfo {title} {Generalization of the
  {M}ulliken-{H}ush treatment for the calculation of electron transfer matrix
  elements},\ }\href {https://doi.org/10.1016/0009-2614(95)01310-5} {\bibfield
  {journal} {\bibinfo  {journal} {Chem. Phys. Lett.}\ }\textbf {\bibinfo
  {volume} {249}},\ \bibinfo {pages} {15} (\bibinfo {year} {1996})}\BibitemShut
  {NoStop}%
\bibitem [{\citenamefont {Dobbyn}\ and\ \citenamefont
  {Knowles}(1997)}]{dobbyn1997}%
  \BibitemOpen
  \bibfield  {author} {\bibinfo {author} {\bibfnamefont {A.~J.}\ \bibnamefont
  {Dobbyn}}\ and\ \bibinfo {author} {\bibfnamefont {P.~J.}\ \bibnamefont
  {Knowles}},\ }\bibfield  {title} {\bibinfo {title} {A comparative study of
  methods for describing non-adiabatic coupling: diabatic representation of the
  1sigma+/1pi {HOH} and {HHO} conical intersections},\ }\href
  {https://doi.org/10.1080/002689797170842} {\bibfield  {journal} {\bibinfo
  {journal} {Mol. Phys.}\ }\textbf {\bibinfo {volume} {91}},\ \bibinfo {pages}
  {1107} (\bibinfo {year} {1997})}\BibitemShut {NoStop}%
\bibitem [{\citenamefont {Evenhuis}\ and\ \citenamefont
  {Collins}(2004)}]{evenhuis2004}%
  \BibitemOpen
  \bibfield  {author} {\bibinfo {author} {\bibfnamefont {C.~R.}\ \bibnamefont
  {Evenhuis}}\ and\ \bibinfo {author} {\bibfnamefont {M.~A.}\ \bibnamefont
  {Collins}},\ }\bibfield  {title} {\bibinfo {title} {Interpolation of diabatic
  potential energy surfaces},\ }\href {https://doi.org/10.1063/1.1770756}
  {\bibfield  {journal} {\bibinfo  {journal} {J. Chem. Phys.}\ }\textbf
  {\bibinfo {volume} {121}},\ \bibinfo {pages} {2515} (\bibinfo {year}
  {2004})}\BibitemShut {NoStop}%
\bibitem [{\citenamefont {Yarkony}\ \emph {et~al.}(2019)\citenamefont
  {Yarkony}, \citenamefont {Xie}, \citenamefont {Zhu}, \citenamefont {Wang},
  \citenamefont {Malbon},\ and\ \citenamefont {Guo}}]{yarkony2019}%
  \BibitemOpen
  \bibfield  {author} {\bibinfo {author} {\bibfnamefont {D.~R.}\ \bibnamefont
  {Yarkony}}, \bibinfo {author} {\bibfnamefont {C.}~\bibnamefont {Xie}},
  \bibinfo {author} {\bibfnamefont {X.}~\bibnamefont {Zhu}}, \bibinfo {author}
  {\bibfnamefont {Y.}~\bibnamefont {Wang}}, \bibinfo {author} {\bibfnamefont
  {C.~L.}\ \bibnamefont {Malbon}},\ and\ \bibinfo {author} {\bibfnamefont
  {H.}~\bibnamefont {Guo}},\ }\bibfield  {title} {\bibinfo {title} {Diabatic
  and adiabatic representations: Electronic structure caveats},\ }\href
  {https://doi.org/10.1016/j.comptc.2019.01.020} {\bibfield  {journal}
  {\bibinfo  {journal} {Comput. Theor. Chem.}\ }\textbf {\bibinfo {volume}
  {1152}},\ \bibinfo {pages} {41} (\bibinfo {year} {2019})}\BibitemShut
  {NoStop}%
\bibitem [{\citenamefont {Wang}\ \emph {et~al.}(2019)\citenamefont {Wang},
  \citenamefont {Guan},\ and\ \citenamefont {Yarkony}}]{wang2019}%
  \BibitemOpen
  \bibfield  {author} {\bibinfo {author} {\bibfnamefont {Y.}~\bibnamefont
  {Wang}}, \bibinfo {author} {\bibfnamefont {Y.}~\bibnamefont {Guan}},\ and\
  \bibinfo {author} {\bibfnamefont {D.~R.}\ \bibnamefont {Yarkony}},\
  }\bibfield  {title} {\bibinfo {title} {On the impact of singularities in the
  two-state adiabatic to diabatic state transformation: A global treatment},\
  }\href {https://doi.org/10.1021/acs.jpca.9b08519} {\bibfield  {journal}
  {\bibinfo  {journal} {J. Chem. Phys. A}\ }\textbf {\bibinfo {volume} {123}},\
  \bibinfo {pages} {9874} (\bibinfo {year} {2019})}\BibitemShut {NoStop}%
\bibitem [{\citenamefont {Han}\ \emph {et~al.}(2020)\citenamefont {Han},
  \citenamefont {Wang}, \citenamefont {Guan}, \citenamefont {Yarkony},\ and\
  \citenamefont {Guo}}]{han2020}%
  \BibitemOpen
  \bibfield  {author} {\bibinfo {author} {\bibfnamefont {S.}~\bibnamefont
  {Han}}, \bibinfo {author} {\bibfnamefont {Y.}~\bibnamefont {Wang}}, \bibinfo
  {author} {\bibfnamefont {Y.}~\bibnamefont {Guan}}, \bibinfo {author}
  {\bibfnamefont {D.~R.}\ \bibnamefont {Yarkony}},\ and\ \bibinfo {author}
  {\bibfnamefont {H.}~\bibnamefont {Guo}},\ }\bibfield  {title} {\bibinfo
  {title} {Impact of diabolical singular points on nonadiabatic dynamics and a
  remedy: Photodissociation of ammonia in the first band},\ }\href
  {https://doi.org/10.1021/acs.jctc.0c00811} {\bibfield  {journal} {\bibinfo
  {journal} {J. Chem. Theory Comput.}\ }\textbf {\bibinfo {volume} {16}},\
  \bibinfo {pages} {6776} (\bibinfo {year} {2020})}\BibitemShut {NoStop}%
\bibitem [{\citenamefont {Li}\ \emph {et~al.}(2023)\citenamefont {Li},
  \citenamefont {Hou},\ and\ \citenamefont {Xie}}]{li2023}%
  \BibitemOpen
  \bibfield  {author} {\bibinfo {author} {\bibfnamefont {C.}~\bibnamefont
  {Li}}, \bibinfo {author} {\bibfnamefont {S.}~\bibnamefont {Hou}},\ and\
  \bibinfo {author} {\bibfnamefont {C.}~\bibnamefont {Xie}},\ }\bibfield
  {title} {\bibinfo {title} {Constructing diabatic potential energy matrices
  with neural networks based on adiabatic energies and physical considerations:
  Toward quantum dynamic accuracy},\ }\href
  {https://doi.org/10.1021/acs.jctc.2c01074} {\bibfield  {journal} {\bibinfo
  {journal} {J. Chem. Theory Comput.}\ }\textbf {\bibinfo {volume} {19}},\
  \bibinfo {pages} {3063} (\bibinfo {year} {2023})}\BibitemShut {NoStop}%
\bibitem [{\citenamefont {Wang}\ \emph
  {et~al.}(2023{\natexlab{a}})\citenamefont {Wang}, \citenamefont {An},
  \citenamefont {Chen}, \citenamefont {Hu}, \citenamefont {Guo},\ and\
  \citenamefont {Xie}}]{wang2023.2}%
  \BibitemOpen
  \bibfield  {author} {\bibinfo {author} {\bibfnamefont {J.}~\bibnamefont
  {Wang}}, \bibinfo {author} {\bibfnamefont {F.}~\bibnamefont {An}}, \bibinfo
  {author} {\bibfnamefont {J.}~\bibnamefont {Chen}}, \bibinfo {author}
  {\bibfnamefont {X.}~\bibnamefont {Hu}}, \bibinfo {author} {\bibfnamefont
  {H.}~\bibnamefont {Guo}},\ and\ \bibinfo {author} {\bibfnamefont
  {D.}~\bibnamefont {Xie}},\ }\bibfield  {title} {\bibinfo {title} {Accurate
  full-dimensional global diabatic potential energy matrix for the two
  lowest-lying electronic states of the {H}+ {O$_{2}$} {$\rightarrow$} {HO} +
  {O} reaction},\ }\href {https://doi.org/10.1021/acs.jctc.3c00291} {\bibfield
  {journal} {\bibinfo  {journal} {J. Chem. Theory Comput.}\ }\textbf {\bibinfo
  {volume} {19}},\ \bibinfo {pages} {2929} (\bibinfo {year}
  {2023}{\natexlab{a}})}\BibitemShut {NoStop}%
\bibitem [{\citenamefont {Vandaele}\ \emph {et~al.}(2024)\citenamefont
  {Vandaele}, \citenamefont {Mali\u{s}},\ and\ \citenamefont
  {Luber}}]{vandaele2024}%
  \BibitemOpen
  \bibfield  {author} {\bibinfo {author} {\bibfnamefont {E.}~\bibnamefont
  {Vandaele}}, \bibinfo {author} {\bibfnamefont {M.}~\bibnamefont
  {Mali\u{s}}},\ and\ \bibinfo {author} {\bibfnamefont {S.}~\bibnamefont
  {Luber}},\ }\bibfield  {title} {\bibinfo {title} {A local diabatisation
  method for two-state adiabatic conical intersections},\ }\href
  {https://doi.org/10.1021/acs.jctc.3c01008} {\bibfield  {journal} {\bibinfo
  {journal} {J. Chem. Theory Comput.}\ }\textbf {\bibinfo {volume} {20}},\
  \bibinfo {pages} {856} (\bibinfo {year} {2024})}\BibitemShut {NoStop}%
\bibitem [{\citenamefont {Ollitrault}\ \emph {et~al.}(2020)\citenamefont
  {Ollitrault}, \citenamefont {Mazzola},\ and\ \citenamefont
  {Tavernelli}}]{ollitrault2020}%
  \BibitemOpen
  \bibfield  {author} {\bibinfo {author} {\bibfnamefont {P.~J.}\ \bibnamefont
  {Ollitrault}}, \bibinfo {author} {\bibfnamefont {G.}~\bibnamefont
  {Mazzola}},\ and\ \bibinfo {author} {\bibfnamefont {I.}~\bibnamefont
  {Tavernelli}},\ }\bibfield  {title} {\bibinfo {title} {Nonadiabatic molecular
  quantum dynamics with quantum computers},\ }\href
  {https://doi.org/10.1103/PhysRevLett.125.260511} {\bibfield  {journal}
  {\bibinfo  {journal} {Phys. Rev. Lett.}\ }\textbf {\bibinfo {volume} {125}},\
  \bibinfo {pages} {260511} (\bibinfo {year} {2020})}\BibitemShut {NoStop}%
\bibitem [{\citenamefont {Ollitrault}\ \emph {et~al.}(2021)\citenamefont
  {Ollitrault}, \citenamefont {Miessen},\ and\ \citenamefont
  {Tavernelli}}]{ollitrault2021}%
  \BibitemOpen
  \bibfield  {author} {\bibinfo {author} {\bibfnamefont {P.~J.}\ \bibnamefont
  {Ollitrault}}, \bibinfo {author} {\bibfnamefont {A.}~\bibnamefont
  {Miessen}},\ and\ \bibinfo {author} {\bibfnamefont {I.}~\bibnamefont
  {Tavernelli}},\ }\bibfield  {title} {\bibinfo {title} {Molecular quantum
  dynamics: A quantum computing perspective},\ }\href
  {https://doi.org/10.1021/acs.accounts.1c00514} {\bibfield  {journal}
  {\bibinfo  {journal} {Acc. Chem. Res.}\ }\textbf {\bibinfo {volume} {54}},\
  \bibinfo {pages} {4229} (\bibinfo {year} {2021})}\BibitemShut {NoStop}%
\bibitem [{\citenamefont {Whitlow}\ \emph {et~al.}(2023)\citenamefont
  {Whitlow}, \citenamefont {Jia}, \citenamefont {Wang}, \citenamefont {Fang},
  \citenamefont {Kim},\ and\ \citenamefont {Brown}}]{whitlow2023quantum}%
  \BibitemOpen
  \bibfield  {author} {\bibinfo {author} {\bibfnamefont {J.}~\bibnamefont
  {Whitlow}}, \bibinfo {author} {\bibfnamefont {Z.}~\bibnamefont {Jia}},
  \bibinfo {author} {\bibfnamefont {Y.}~\bibnamefont {Wang}}, \bibinfo {author}
  {\bibfnamefont {C.}~\bibnamefont {Fang}}, \bibinfo {author} {\bibfnamefont
  {J.}~\bibnamefont {Kim}},\ and\ \bibinfo {author} {\bibfnamefont {K.~R.}\
  \bibnamefont {Brown}},\ }\bibfield  {title} {\bibinfo {title} {Quantum
  simulation of conical intersections using trapped ions},\ }\href
  {https://doi.org/10.1038/s41557-023-01303-0} {\bibfield  {journal} {\bibinfo
  {journal} {Nat. Chem.}\ }\textbf {\bibinfo {volume} {15}},\ \bibinfo {pages}
  {1509} (\bibinfo {year} {2023})}\BibitemShut {NoStop}%
\bibitem [{\citenamefont {Valahu}\ \emph {et~al.}(2023)\citenamefont {Valahu},
  \citenamefont {Olaya-Agudelo}, \citenamefont {MacDonell}, \citenamefont
  {Navickas}, \citenamefont {Rao}, \citenamefont {Millican}, \citenamefont
  {Pérez-Sánchez}, \citenamefont {Yuen-Zhou}, \citenamefont {Biercuk},
  \citenamefont {Hempel}, \citenamefont {Tan},\ and\ \citenamefont
  {Kassal}}]{Valahu.2023}%
  \BibitemOpen
  \bibfield  {author} {\bibinfo {author} {\bibfnamefont {C.~H.}\ \bibnamefont
  {Valahu}}, \bibinfo {author} {\bibfnamefont {V.~C.}\ \bibnamefont
  {Olaya-Agudelo}}, \bibinfo {author} {\bibfnamefont {R.~J.}\ \bibnamefont
  {MacDonell}}, \bibinfo {author} {\bibfnamefont {T.}~\bibnamefont {Navickas}},
  \bibinfo {author} {\bibfnamefont {A.~D.}\ \bibnamefont {Rao}}, \bibinfo
  {author} {\bibfnamefont {M.~J.}\ \bibnamefont {Millican}}, \bibinfo {author}
  {\bibfnamefont {J.~B.}\ \bibnamefont {Pérez-Sánchez}}, \bibinfo {author}
  {\bibfnamefont {J.}~\bibnamefont {Yuen-Zhou}}, \bibinfo {author}
  {\bibfnamefont {M.~J.}\ \bibnamefont {Biercuk}}, \bibinfo {author}
  {\bibfnamefont {C.}~\bibnamefont {Hempel}}, \bibinfo {author} {\bibfnamefont
  {T.~R.}\ \bibnamefont {Tan}},\ and\ \bibinfo {author} {\bibfnamefont
  {I.}~\bibnamefont {Kassal}},\ }\bibfield  {title} {\bibinfo {title} {{Direct
  observation of geometric-phase interference in dynamics around a conical
  intersection}},\ }\href {https://doi.org/10.1038/s41557-023-01300-3}
  {\bibfield  {journal} {\bibinfo  {journal} {Nat. Chem.}\ }\textbf {\bibinfo
  {volume} {15}},\ \bibinfo {pages} {1503} (\bibinfo {year} {2023})},\ \Eprint
  {https://arxiv.org/abs/2211.07320} {2211.07320} \BibitemShut {NoStop}%
\bibitem [{\citenamefont {Wang}\ \emph
  {et~al.}(2023{\natexlab{b}})\citenamefont {Wang}, \citenamefont {Frattini},
  \citenamefont {Chapman}, \citenamefont {Puri}, \citenamefont {Girvin},
  \citenamefont {Devoret},\ and\ \citenamefont {Schoelkopf}}]{Wang.2023hrm}%
  \BibitemOpen
  \bibfield  {author} {\bibinfo {author} {\bibfnamefont {C.~S.}\ \bibnamefont
  {Wang}}, \bibinfo {author} {\bibfnamefont {N.~E.}\ \bibnamefont {Frattini}},
  \bibinfo {author} {\bibfnamefont {B.~J.}\ \bibnamefont {Chapman}}, \bibinfo
  {author} {\bibfnamefont {S.}~\bibnamefont {Puri}}, \bibinfo {author}
  {\bibfnamefont {S.~M.}\ \bibnamefont {Girvin}}, \bibinfo {author}
  {\bibfnamefont {M.~H.}\ \bibnamefont {Devoret}},\ and\ \bibinfo {author}
  {\bibfnamefont {R.~J.}\ \bibnamefont {Schoelkopf}},\ }\bibfield  {title}
  {\bibinfo {title} {{Observation of Wave-Packet Branching through an
  Engineered Conical Intersection}},\ }\href
  {https://doi.org/10.1103/physrevx.13.011008} {\bibfield  {journal} {\bibinfo
  {journal} {Phys. Rev. X}\ }\textbf {\bibinfo {volume} {13}},\ \bibinfo
  {pages} {011008} (\bibinfo {year} {2023}{\natexlab{b}})},\ \Eprint
  {https://arxiv.org/abs/2202.02364} {2202.02364} \BibitemShut {NoStop}%
\bibitem [{\citenamefont {Anand}\ \emph {et~al.}(2022)\citenamefont {Anand},
  \citenamefont {Schleich}, \citenamefont {Alperin-Lea}, \citenamefont
  {Jensen}, \citenamefont {Sim}, \citenamefont {D{\'\i}az-Tinoco},
  \citenamefont {Kottmann}, \citenamefont {Degroote}, \citenamefont
  {Izmaylov},\ and\ \citenamefont {Aspuru-Guzik}}]{anand2022}%
  \BibitemOpen
  \bibfield  {author} {\bibinfo {author} {\bibfnamefont {A.}~\bibnamefont
  {Anand}}, \bibinfo {author} {\bibfnamefont {P.}~\bibnamefont {Schleich}},
  \bibinfo {author} {\bibfnamefont {S.}~\bibnamefont {Alperin-Lea}}, \bibinfo
  {author} {\bibfnamefont {P.~W.}\ \bibnamefont {Jensen}}, \bibinfo {author}
  {\bibfnamefont {S.}~\bibnamefont {Sim}}, \bibinfo {author} {\bibfnamefont
  {M.}~\bibnamefont {D{\'\i}az-Tinoco}}, \bibinfo {author} {\bibfnamefont
  {J.~S.}\ \bibnamefont {Kottmann}}, \bibinfo {author} {\bibfnamefont
  {M.}~\bibnamefont {Degroote}}, \bibinfo {author} {\bibfnamefont {A.~F.}\
  \bibnamefont {Izmaylov}},\ and\ \bibinfo {author} {\bibfnamefont
  {A.}~\bibnamefont {Aspuru-Guzik}},\ }\bibfield  {title} {\bibinfo {title} {A
  quantum computing view on unitary coupled cluster theory},\ }\href
  {https://doi.org/10.1039/D1CS00932J} {\bibfield  {journal} {\bibinfo
  {journal} {Chem. Soc. Rev.}\ }\textbf {\bibinfo {volume} {51}},\ \bibinfo
  {pages} {1659} (\bibinfo {year} {2022})}\BibitemShut {NoStop}%
\bibitem [{\citenamefont {Lee}\ \emph {et~al.}(2018)\citenamefont {Lee},
  \citenamefont {Huggins}, \citenamefont {Head-Gordon},\ and\ \citenamefont
  {Whaley}}]{lee2018}%
  \BibitemOpen
  \bibfield  {author} {\bibinfo {author} {\bibfnamefont {J.}~\bibnamefont
  {Lee}}, \bibinfo {author} {\bibfnamefont {W.~J.}\ \bibnamefont {Huggins}},
  \bibinfo {author} {\bibfnamefont {M.}~\bibnamefont {Head-Gordon}},\ and\
  \bibinfo {author} {\bibfnamefont {K.~B.}\ \bibnamefont {Whaley}},\ }\bibfield
   {title} {\bibinfo {title} {Generalized unitary coupled cluster wave
  functions for quantum computation},\ }\href
  {https://doi.org/10.1021/acs.jctc.8b01004} {\bibfield  {journal} {\bibinfo
  {journal} {J. Chem. Theory Comput.}\ }\textbf {\bibinfo {volume} {15}},\
  \bibinfo {pages} {311} (\bibinfo {year} {2018})}\BibitemShut {NoStop}%
\bibitem [{\citenamefont {Mazziotti}(2006)}]{mazziotti2006}%
  \BibitemOpen
  \bibfield  {author} {\bibinfo {author} {\bibfnamefont {D.~A.}\ \bibnamefont
  {Mazziotti}},\ }\bibfield  {title} {\bibinfo {title} {Anti-hermitian
  contracted {S}chr{\"o}dinger equation: Direct determination of the
  two-electron reduced density matrices of many-electron molecules},\ }\href
  {https://doi.org/10.1103/PhysRevLett.97.143002} {\bibfield  {journal}
  {\bibinfo  {journal} {Phys. Rev. Lett.}\ }\textbf {\bibinfo {volume} {97}},\
  \bibinfo {pages} {143002} (\bibinfo {year} {2006})}\BibitemShut {NoStop}%
\bibitem [{\citenamefont {Mazziotti}(2007{\natexlab{a}})}]{mazziotti2007}%
  \BibitemOpen
  \bibfield  {author} {\bibinfo {author} {\bibfnamefont {D.~A.}\ \bibnamefont
  {Mazziotti}},\ }\bibfield  {title} {\bibinfo {title} {Anti-hermitian part of
  the contracted {S}chr{\"o}dinger equation for the direct calculation of
  two-electron reduced density matrices},\ }\href
  {https://doi.org/10.1103/PhysRevA.75.022505} {\bibfield  {journal} {\bibinfo
  {journal} {Phys. Rev. A}\ }\textbf {\bibinfo {volume} {75}},\ \bibinfo
  {pages} {022505} (\bibinfo {year} {2007}{\natexlab{a}})}\BibitemShut
  {NoStop}%
\bibitem [{\citenamefont {Mazziotti}(2007{\natexlab{b}})}]{mazziotti2007.2}%
  \BibitemOpen
  \bibfield  {author} {\bibinfo {author} {\bibfnamefont {D.~A.}\ \bibnamefont
  {Mazziotti}},\ }\bibfield  {title} {\bibinfo {title} {Multireference
  many-electron correlation energies from two-electron reduced density matrices
  computed by solving the anti-hermitian contracted {S}chr{\"o}dinger
  equation},\ }\href {https://doi.org/10.1103/PhysRevA.76.052502} {\bibfield
  {journal} {\bibinfo  {journal} {Phys. Rev. A}\ }\textbf {\bibinfo {volume}
  {76}},\ \bibinfo {pages} {052502} (\bibinfo {year}
  {2007}{\natexlab{b}})}\BibitemShut {NoStop}%
\bibitem [{\citenamefont {Smart}\ and\ \citenamefont
  {Mazziotti}(2021)}]{smart2021}%
  \BibitemOpen
  \bibfield  {author} {\bibinfo {author} {\bibfnamefont {S.~E.}\ \bibnamefont
  {Smart}}\ and\ \bibinfo {author} {\bibfnamefont {D.~A.}\ \bibnamefont
  {Mazziotti}},\ }\bibfield  {title} {\bibinfo {title} {Quantum solver of
  contracted eigenvalue equations for scalable molecular simulations on quantum
  computing devices},\ }\href {https://doi.org/10.1103/PhysRevLett.126.070504}
  {\bibfield  {journal} {\bibinfo  {journal} {Phys. Rev. Lett.}\ }\textbf
  {\bibinfo {volume} {126}},\ \bibinfo {pages} {070504} (\bibinfo {year}
  {2021})}\BibitemShut {NoStop}%
\bibitem [{\citenamefont {Smart}\ and\ \citenamefont
  {Mazziotti}(2022)}]{smart2022}%
  \BibitemOpen
  \bibfield  {author} {\bibinfo {author} {\bibfnamefont {S.~E.}\ \bibnamefont
  {Smart}}\ and\ \bibinfo {author} {\bibfnamefont {D.~A.}\ \bibnamefont
  {Mazziotti}},\ }\bibfield  {title} {\bibinfo {title} {Accelerated convergence
  of contracted quantum eigensolvers through a quasi-second-order, locally
  parameterized optimization},\ }\href
  {https://doi.org/10.1021/acs.jctc.2c00446} {\bibfield  {journal} {\bibinfo
  {journal} {J. Chem. Theory Comput.}\ }\textbf {\bibinfo {volume} {18}},\
  \bibinfo {pages} {5286} (\bibinfo {year} {2022})}\BibitemShut {NoStop}%
\bibitem [{\citenamefont {Boyn}\ and\ \citenamefont
  {Mazziotti}(2021)}]{Boyn.2021}%
  \BibitemOpen
  \bibfield  {author} {\bibinfo {author} {\bibfnamefont {J.-N.}\ \bibnamefont
  {Boyn}}\ and\ \bibinfo {author} {\bibfnamefont {D.~A.}\ \bibnamefont
  {Mazziotti}},\ }\bibfield  {title} {\bibinfo {title} {{Accurate
  singlet–triplet gaps in biradicals via the spin averaged anti-Hermitian
  contracted Schrödinger equation}},\ }\href
  {https://doi.org/10.1063/5.0045007} {\bibfield  {journal} {\bibinfo
  {journal} {J. Chem. Phys.}\ }\textbf {\bibinfo {volume} {154}},\ \bibinfo
  {pages} {134103} (\bibinfo {year} {2021})},\ \Eprint
  {https://arxiv.org/abs/2104.00626} {2104.00626} \BibitemShut {NoStop}%
\bibitem [{\citenamefont {Boyn}\ \emph {et~al.}(2021)\citenamefont {Boyn},
  \citenamefont {Lykhin}, \citenamefont {Smart}, \citenamefont {Gagliardi},\
  and\ \citenamefont {Mazziotti}}]{Boyn.2021u94}%
  \BibitemOpen
  \bibfield  {author} {\bibinfo {author} {\bibfnamefont {J.-N.}\ \bibnamefont
  {Boyn}}, \bibinfo {author} {\bibfnamefont {A.~O.}\ \bibnamefont {Lykhin}},
  \bibinfo {author} {\bibfnamefont {S.~E.}\ \bibnamefont {Smart}}, \bibinfo
  {author} {\bibfnamefont {L.}~\bibnamefont {Gagliardi}},\ and\ \bibinfo
  {author} {\bibfnamefont {D.~A.}\ \bibnamefont {Mazziotti}},\ }\bibfield
  {title} {\bibinfo {title} {{Quantum-classical hybrid algorithm for the
  simulation of all-electron correlation}},\ }\href
  {https://doi.org/10.1063/5.0074842} {\bibfield  {journal} {\bibinfo
  {journal} {J. Chem. Phys.}\ }\textbf {\bibinfo {volume} {155}},\ \bibinfo
  {pages} {244106} (\bibinfo {year} {2021})},\ \Eprint
  {https://arxiv.org/abs/2106.11972} {2106.11972} \BibitemShut {NoStop}%
\bibitem [{\citenamefont {Smart}\ \emph {et~al.}(2022)\citenamefont {Smart},
  \citenamefont {Boyn},\ and\ \citenamefont {Mazziotti}}]{Smart.2022w8u}%
  \BibitemOpen
  \bibfield  {author} {\bibinfo {author} {\bibfnamefont {S.~E.}\ \bibnamefont
  {Smart}}, \bibinfo {author} {\bibfnamefont {J.-N.}\ \bibnamefont {Boyn}},\
  and\ \bibinfo {author} {\bibfnamefont {D.~A.}\ \bibnamefont {Mazziotti}},\
  }\bibfield  {title} {\bibinfo {title} {{Resolving correlated states of
  benzyne with an error-mitigated contracted quantum eigensolver}},\ }\href
  {https://doi.org/10.1103/physreva.105.022405} {\bibfield  {journal} {\bibinfo
   {journal} {Phys. Rev. A}\ }\textbf {\bibinfo {volume} {105}},\ \bibinfo
  {pages} {022405} (\bibinfo {year} {2022})},\ \Eprint
  {https://arxiv.org/abs/2103.06876} {2103.06876} \BibitemShut {NoStop}%
\bibitem [{\citenamefont {Wang}\ \emph
  {et~al.}(2023{\natexlab{c}})\citenamefont {Wang}, \citenamefont
  {Sager-Smith},\ and\ \citenamefont {Mazziotti}}]{Wang.2023i8g}%
  \BibitemOpen
  \bibfield  {author} {\bibinfo {author} {\bibfnamefont {Y.}~\bibnamefont
  {Wang}}, \bibinfo {author} {\bibfnamefont {L.~M.}\ \bibnamefont
  {Sager-Smith}},\ and\ \bibinfo {author} {\bibfnamefont {D.~A.}\ \bibnamefont
  {Mazziotti}},\ }\bibfield  {title} {\bibinfo {title} {{Quantum simulation of
  bosons with the contracted quantum eigensolver}},\ }\href
  {https://doi.org/10.1088/1367-2630/acf9c3} {\bibfield  {journal} {\bibinfo
  {journal} {New J. Phys.}\ }\textbf {\bibinfo {volume} {25}},\ \bibinfo
  {pages} {103005} (\bibinfo {year} {2023}{\natexlab{c}})}\BibitemShut
  {NoStop}%
\bibitem [{\citenamefont {Snyder}\ \emph {et~al.}(2010)\citenamefont {Snyder},
  \citenamefont {Rothman}, \citenamefont {Foley},\ and\ \citenamefont
  {Mazziotti}}]{Snyder.2010}%
  \BibitemOpen
  \bibfield  {author} {\bibinfo {author} {\bibfnamefont {J.~W.}\ \bibnamefont
  {Snyder}}, \bibinfo {author} {\bibfnamefont {A.~E.}\ \bibnamefont {Rothman}},
  \bibinfo {author} {\bibfnamefont {J.~J.}\ \bibnamefont {Foley}},\ and\
  \bibinfo {author} {\bibfnamefont {D.~A.}\ \bibnamefont {Mazziotti}},\
  }\bibfield  {title} {\bibinfo {title} {{Conical intersections in triplet
  excited states of methylene from the anti-Hermitian contracted Schrödinger
  equation}},\ }\href {https://doi.org/10.1063/1.3394020} {\bibfield  {journal}
  {\bibinfo  {journal} {J. Chem. Phys.}\ }\textbf {\bibinfo {volume} {132}},\
  \bibinfo {pages} {154109} (\bibinfo {year} {2010})}\BibitemShut {NoStop}%
\bibitem [{\citenamefont {Snyder}\ and\ \citenamefont
  {Mazziotti}(2011{\natexlab{a}})}]{Snyder.2011u3}%
  \BibitemOpen
  \bibfield  {author} {\bibinfo {author} {\bibfnamefont {J.~W.}\ \bibnamefont
  {Snyder}}\ and\ \bibinfo {author} {\bibfnamefont {D.~A.}\ \bibnamefont
  {Mazziotti}},\ }\bibfield  {title} {\bibinfo {title} {{Photoexcited
  conversion of gauche-1,3-butadiene to bicyclobutane via a conical
  intersection: Energies and reduced density matrices from the anti-Hermitian
  contracted Schrödinger equation}},\ }\href
  {https://doi.org/10.1063/1.3606466} {\bibfield  {journal} {\bibinfo
  {journal} {J. Chem. Phys.}\ }\textbf {\bibinfo {volume} {135}},\ \bibinfo
  {pages} {024107} (\bibinfo {year} {2011}{\natexlab{a}})}\BibitemShut
  {NoStop}%
\bibitem [{\citenamefont {Snyder}\ and\ \citenamefont
  {Mazziotti}(2011{\natexlab{b}})}]{Snyder.2011}%
  \BibitemOpen
  \bibfield  {author} {\bibinfo {author} {\bibfnamefont {J.~W.}\ \bibnamefont
  {Snyder}}\ and\ \bibinfo {author} {\bibfnamefont {D.~A.}\ \bibnamefont
  {Mazziotti}},\ }\bibfield  {title} {\bibinfo {title} {{Conical Intersection
  of the Ground and First Excited States of Water: Energies and Reduced Density
  Matrices from the Anti-Hermitian Contracted Schr{\"o}dinger Equation}},\
  }\href {https://doi.org/10.1021/jp208013m} {\bibfield  {journal} {\bibinfo
  {journal} {J. Phys. Chem. A}\ }\textbf {\bibinfo {volume} {115}},\ \bibinfo
  {pages} {14120} (\bibinfo {year} {2011}{\natexlab{b}})}\BibitemShut {NoStop}%
\bibitem [{\citenamefont {Snyder}\ and\ \citenamefont
  {Mazziotti}(2011{\natexlab{c}})}]{Snyder.2011b}%
  \BibitemOpen
  \bibfield  {author} {\bibinfo {author} {\bibfnamefont {J.~W.}\ \bibnamefont
  {Snyder}}\ and\ \bibinfo {author} {\bibfnamefont {D.~A.}\ \bibnamefont
  {Mazziotti}},\ }\bibfield  {title} {\bibinfo {title} {{Photoexcited
  tautomerization of vinyl alcohol to acetylaldehyde via a conical intersection
  from contracted Schrödinger theory}},\ }\href
  {https://doi.org/10.1039/c2cp23065h} {\bibfield  {journal} {\bibinfo
  {journal} {Phys. Chem. Chem. Phys.}\ }\textbf {\bibinfo {volume} {14}},\
  \bibinfo {pages} {1660} (\bibinfo {year} {2011}{\natexlab{c}})}\BibitemShut
  {NoStop}%
\bibitem [{\citenamefont {Schlimgen}\ \emph {et~al.}(2021)\citenamefont
  {Schlimgen}, \citenamefont {Head-Marsden}, \citenamefont {Sager},
  \citenamefont {Narang},\ and\ \citenamefont {Mazziotti}}]{schlimgen2021}%
  \BibitemOpen
  \bibfield  {author} {\bibinfo {author} {\bibfnamefont {A.~W.}\ \bibnamefont
  {Schlimgen}}, \bibinfo {author} {\bibfnamefont {K.}~\bibnamefont
  {Head-Marsden}}, \bibinfo {author} {\bibfnamefont {L.~M.}\ \bibnamefont
  {Sager}}, \bibinfo {author} {\bibfnamefont {P.}~\bibnamefont {Narang}},\ and\
  \bibinfo {author} {\bibfnamefont {D.~A.}\ \bibnamefont {Mazziotti}},\
  }\bibfield  {title} {\bibinfo {title} {Quantum simulation of open quantum
  systems using a unitary decomposition of operators},\ }\href
  {https://doi.org/10.1103/PhysRevLett.127.270503} {\bibfield  {journal}
  {\bibinfo  {journal} {Phys. Rev. Lett.}\ }\textbf {\bibinfo {volume} {127}},\
  \bibinfo {pages} {270503} (\bibinfo {year} {2021})}\BibitemShut {NoStop}%
\bibitem [{\citenamefont {Hu}\ \emph {et~al.}(2020)\citenamefont {Hu},
  \citenamefont {Xia},\ and\ \citenamefont {Kais}}]{Hu.2020}%
  \BibitemOpen
  \bibfield  {author} {\bibinfo {author} {\bibfnamefont {Z.}~\bibnamefont
  {Hu}}, \bibinfo {author} {\bibfnamefont {R.}~\bibnamefont {Xia}},\ and\
  \bibinfo {author} {\bibfnamefont {S.}~\bibnamefont {Kais}},\ }\bibfield
  {title} {\bibinfo {title} {{A quantum algorithm for evolving open quantum
  dynamics on quantum computing devices}},\ }\href
  {https://doi.org/10.1038/s41598-020-60321-x} {\bibfield  {journal} {\bibinfo
  {journal} {Scientific Reports}\ }\textbf {\bibinfo {volume} {10}},\ \bibinfo
  {pages} {3301} (\bibinfo {year} {2020})},\ \Eprint
  {https://arxiv.org/abs/1904.00910} {1904.00910} \BibitemShut {NoStop}%
\bibitem [{\citenamefont {Longuet-Higgins}\ \emph {et~al.}(1958)\citenamefont
  {Longuet-Higgins}, \citenamefont {{\"O}pik}, \citenamefont {Pryce},\ and\
  \citenamefont {Sack}}]{longuet1958}%
  \BibitemOpen
  \bibfield  {author} {\bibinfo {author} {\bibfnamefont {H.~C.}\ \bibnamefont
  {Longuet-Higgins}}, \bibinfo {author} {\bibfnamefont {U.}~\bibnamefont
  {{\"O}pik}}, \bibinfo {author} {\bibfnamefont {M.~H.~L.}\ \bibnamefont
  {Pryce}},\ and\ \bibinfo {author} {\bibfnamefont {R.}~\bibnamefont {Sack}},\
  }\bibfield  {title} {\bibinfo {title} {Studies of the {J}ahn-{T}eller effect.
  ii. the dynamical problem},\ }\href {https://doi.org/10.1098/rspa.1958.0022}
  {\bibfield  {journal} {\bibinfo  {journal} {Proc. R. Soc. Lond. A}\ }\textbf
  {\bibinfo {volume} {244}},\ \bibinfo {pages} {1} (\bibinfo {year}
  {1958})}\BibitemShut {NoStop}%
\bibitem [{\citenamefont {Wang}\ and\ \citenamefont
  {Mazziotti}(2023)}]{wang2023}%
  \BibitemOpen
  \bibfield  {author} {\bibinfo {author} {\bibfnamefont {Y.}~\bibnamefont
  {Wang}}\ and\ \bibinfo {author} {\bibfnamefont {D.~A.}\ \bibnamefont
  {Mazziotti}},\ }\bibfield  {title} {\bibinfo {title} {Electronic excited
  states from a variance-based contracted quantum eigensolver},\ }\href
  {https://doi.org/10.1103/PhysRevA.108.022814} {\bibfield  {journal} {\bibinfo
   {journal} {Phys. Rev. A}\ ,\ \bibinfo {pages} {022814}} (\bibinfo {year}
  {2023})}\BibitemShut {NoStop}%
\bibitem [{\citenamefont {Mead}(1992)}]{mead1992}%
  \BibitemOpen
  \bibfield  {author} {\bibinfo {author} {\bibfnamefont {C.~A.}\ \bibnamefont
  {Mead}},\ }\bibfield  {title} {\bibinfo {title} {The geometric phase in
  molecular systems},\ }\href {https://doi.org/10.1103/RevModPhys.64.51}
  {\bibfield  {journal} {\bibinfo  {journal} {Rev. Mod. Phys.}\ }\textbf
  {\bibinfo {volume} {64}},\ \bibinfo {pages} {51} (\bibinfo {year}
  {1992})}\BibitemShut {NoStop}%
\bibitem [{\citenamefont {Xie}\ \emph {et~al.}(2019)\citenamefont {Xie},
  \citenamefont {Malbon}, \citenamefont {Guo},\ and\ \citenamefont
  {Yarkony}}]{xie2019}%
  \BibitemOpen
  \bibfield  {author} {\bibinfo {author} {\bibfnamefont {C.}~\bibnamefont
  {Xie}}, \bibinfo {author} {\bibfnamefont {C.~L.}\ \bibnamefont {Malbon}},
  \bibinfo {author} {\bibfnamefont {H.}~\bibnamefont {Guo}},\ and\ \bibinfo
  {author} {\bibfnamefont {D.~R.}\ \bibnamefont {Yarkony}},\ }\bibfield
  {title} {\bibinfo {title} {Up to a sign. the insidious effects of
  energetically inaccessible conical intersections on unimolecular reactions},\
  }\href {https://doi.org/10.1021/acs.accounts.8b00571} {\bibfield  {journal}
  {\bibinfo  {journal} {Acc. Chem. Res.}\ }\textbf {\bibinfo {volume} {52}},\
  \bibinfo {pages} {501} (\bibinfo {year} {2019})}\BibitemShut {NoStop}%
\bibitem [{\citenamefont {Wang}\ and\ \citenamefont
  {Yarkony}(2021)}]{wang2021}%
  \BibitemOpen
  \bibfield  {author} {\bibinfo {author} {\bibfnamefont {Y.}~\bibnamefont
  {Wang}}\ and\ \bibinfo {author} {\bibfnamefont {D.~R.}\ \bibnamefont
  {Yarkony}},\ }\bibfield  {title} {\bibinfo {title} {Conical intersection
  seams in spin--orbit coupled systems with an even number of electrons: A
  numerical study based on neural network fit surfaces},\ }\href
  {https://doi.org/10.1063/5.0067660} {\bibfield  {journal} {\bibinfo
  {journal} {J. Chem. Phys.}\ }\textbf {\bibinfo {volume} {155}},\ \bibinfo
  {pages} {174115} (\bibinfo {year} {2021})}\BibitemShut {NoStop}%
\bibitem [{\citenamefont {Kamisaka}\ \emph {et~al.}(2002)\citenamefont
  {Kamisaka}, \citenamefont {Bian}, \citenamefont {Nobusada},\ and\
  \citenamefont {Nakamura}}]{kamisaka2002}%
  \BibitemOpen
  \bibfield  {author} {\bibinfo {author} {\bibfnamefont {H.}~\bibnamefont
  {Kamisaka}}, \bibinfo {author} {\bibfnamefont {W.}~\bibnamefont {Bian}},
  \bibinfo {author} {\bibfnamefont {K.}~\bibnamefont {Nobusada}},\ and\
  \bibinfo {author} {\bibfnamefont {H.}~\bibnamefont {Nakamura}},\ }\bibfield
  {title} {\bibinfo {title} {Accurate quantum dynamics of electronically
  nonadiabatic chemical reactions in the {DH}$_{2}^{+}$ system},\ }\href
  {https://doi.org/10.1063/1.1418252} {\bibfield  {journal} {\bibinfo
  {journal} {J. Chem. Phys.}\ }\textbf {\bibinfo {volume} {116}},\ \bibinfo
  {pages} {654} (\bibinfo {year} {2002})}\BibitemShut {NoStop}%
\bibitem [{\citenamefont {Viegas}\ \emph {et~al.}(2007)\citenamefont {Viegas},
  \citenamefont {Alijah},\ and\ \citenamefont {Varandas}}]{viegas2007}%
  \BibitemOpen
  \bibfield  {author} {\bibinfo {author} {\bibfnamefont {L.~P.}\ \bibnamefont
  {Viegas}}, \bibinfo {author} {\bibfnamefont {A.}~\bibnamefont {Alijah}},\
  and\ \bibinfo {author} {\bibfnamefont {A.~J.}\ \bibnamefont {Varandas}},\
  }\bibfield  {title} {\bibinfo {title} {Accurate ab initio based multisheeted
  double many-body expansion potential energy surface for the three lowest
  electronic singlet states of {H}$_{3}^{+}$},\ }\href
  {https://doi.org/10.1063/1.2566770} {\bibfield  {journal} {\bibinfo
  {journal} {J. Chem. Phys.}\ }\textbf {\bibinfo {volume} {126}},\ \bibinfo
  {pages} {074309} (\bibinfo {year} {2007})}\BibitemShut {NoStop}%
\bibitem [{\citenamefont {Barrag{\'a}n}\ \emph {et~al.}(2006)\citenamefont
  {Barrag{\'a}n}, \citenamefont {Errea}, \citenamefont {Mac{\'\i}as},
  \citenamefont {M{\'e}ndez}, \citenamefont {Rabad{\'a}n},\ and\ \citenamefont
  {Riera}}]{barragan2006}%
  \BibitemOpen
  \bibfield  {author} {\bibinfo {author} {\bibfnamefont {P.}~\bibnamefont
  {Barrag{\'a}n}}, \bibinfo {author} {\bibfnamefont {L.}~\bibnamefont {Errea}},
  \bibinfo {author} {\bibfnamefont {A.}~\bibnamefont {Mac{\'\i}as}}, \bibinfo
  {author} {\bibfnamefont {L.}~\bibnamefont {M{\'e}ndez}}, \bibinfo {author}
  {\bibfnamefont {I.}~\bibnamefont {Rabad{\'a}n}},\ and\ \bibinfo {author}
  {\bibfnamefont {A.}~\bibnamefont {Riera}},\ }\bibfield  {title} {\bibinfo
  {title} {A study of conical intersections for the {H}$_{3}^{+}$ system},\
  }\href {https://doi.org/10.1063/1.2193516} {\bibfield  {journal} {\bibinfo
  {journal} {J. Chem. Phys.}\ }\textbf {\bibinfo {volume} {124}},\ \bibinfo
  {pages} {184303} (\bibinfo {year} {2006})}\BibitemShut {NoStop}%
\bibitem [{\citenamefont {Mukherjee}\ \emph {et~al.}(2014)\citenamefont
  {Mukherjee}, \citenamefont {Mukhopadhyay},\ and\ \citenamefont
  {Adhikari}}]{mukherjee2014}%
  \BibitemOpen
  \bibfield  {author} {\bibinfo {author} {\bibfnamefont {S.}~\bibnamefont
  {Mukherjee}}, \bibinfo {author} {\bibfnamefont {D.}~\bibnamefont
  {Mukhopadhyay}},\ and\ \bibinfo {author} {\bibfnamefont {S.}~\bibnamefont
  {Adhikari}},\ }\bibfield  {title} {\bibinfo {title} {Conical intersections
  and diabatic potential energy surfaces for the three lowest electronic
  singlet states of {H}$_{3}^{+}$},\ }\href {https://doi.org/10.1063/1.4901986}
  {\bibfield  {journal} {\bibinfo  {journal} {J. Chem. Phys.}\ }\textbf
  {\bibinfo {volume} {141}},\ \bibinfo {pages} {204306} (\bibinfo {year}
  {2014})}\BibitemShut {NoStop}%
\bibitem [{\citenamefont {Ghosh}\ \emph {et~al.}(2017)\citenamefont {Ghosh},
  \citenamefont {Mukherjee}, \citenamefont {Mukherjee}, \citenamefont {Mandal},
  \citenamefont {Sharma}, \citenamefont {Chaudhury},\ and\ \citenamefont
  {Adhikari}}]{ghosh2017}%
  \BibitemOpen
  \bibfield  {author} {\bibinfo {author} {\bibfnamefont {S.}~\bibnamefont
  {Ghosh}}, \bibinfo {author} {\bibfnamefont {S.}~\bibnamefont {Mukherjee}},
  \bibinfo {author} {\bibfnamefont {B.}~\bibnamefont {Mukherjee}}, \bibinfo
  {author} {\bibfnamefont {S.}~\bibnamefont {Mandal}}, \bibinfo {author}
  {\bibfnamefont {R.}~\bibnamefont {Sharma}}, \bibinfo {author} {\bibfnamefont
  {P.}~\bibnamefont {Chaudhury}},\ and\ \bibinfo {author} {\bibfnamefont
  {S.}~\bibnamefont {Adhikari}},\ }\bibfield  {title} {\bibinfo {title} {Beyond
  born-oppenheimer theory for ab initio constructed diabatic potential energy
  surfaces of singlet {H}$_{3}^{+}$ to study reaction dynamics using coupled 3d
  time-dependent wave-packet approach},\ }\href
  {https://doi.org/10.1063/1.4998406} {\bibfield  {journal} {\bibinfo
  {journal} {J. Chem. Phys.}\ }\textbf {\bibinfo {volume} {147}},\ \bibinfo
  {pages} {074105} (\bibinfo {year} {2017})}\BibitemShut {NoStop}%
\bibitem [{\citenamefont {Yin}\ \emph {et~al.}(2021)\citenamefont {Yin},
  \citenamefont {Braams}, \citenamefont {Fu},\ and\ \citenamefont
  {Zhang}}]{yin2021}%
  \BibitemOpen
  \bibfield  {author} {\bibinfo {author} {\bibfnamefont {Z.}~\bibnamefont
  {Yin}}, \bibinfo {author} {\bibfnamefont {B.~J.}\ \bibnamefont {Braams}},
  \bibinfo {author} {\bibfnamefont {B.}~\bibnamefont {Fu}},\ and\ \bibinfo
  {author} {\bibfnamefont {D.~H.}\ \bibnamefont {Zhang}},\ }\bibfield  {title}
  {\bibinfo {title} {Neural network representation of three-state quasidiabatic
  hamiltonians based on the transformation properties from a valence bond
  model: Three singlet states of {H}$_{3}^{+}$},\ }\href
  {https://doi.org/10.1021/acs.jctc.0c01336} {\bibfield  {journal} {\bibinfo
  {journal} {J. Chem. Theory Comput.}\ }\textbf {\bibinfo {volume} {17}},\
  \bibinfo {pages} {1678} (\bibinfo {year} {2021})}\BibitemShut {NoStop}%
\bibitem [{\citenamefont {Guan}\ \emph {et~al.}(2022)\citenamefont {Guan},
  \citenamefont {Yarkony},\ and\ \citenamefont {Zhang}}]{guan2022}%
  \BibitemOpen
  \bibfield  {author} {\bibinfo {author} {\bibfnamefont {Y.}~\bibnamefont
  {Guan}}, \bibinfo {author} {\bibfnamefont {D.~R.}\ \bibnamefont {Yarkony}},\
  and\ \bibinfo {author} {\bibfnamefont {D.~H.}\ \bibnamefont {Zhang}},\
  }\bibfield  {title} {\bibinfo {title} {Permutation invariant polynomial
  neural network based diabatic ansatz for the ({E}+ {A})$\times$(e+ a)
  {J}ahn--{T}eller and pseudo-{J}ahn--{T}eller systems},\ }\href
  {https://doi.org/10.1063/5.0096912} {\bibfield  {journal} {\bibinfo
  {journal} {J. Chem. Phys.}\ }\textbf {\bibinfo {volume} {157}},\ \bibinfo
  {pages} {014110} (\bibinfo {year} {2022})}\BibitemShut {NoStop}%
\bibitem [{\citenamefont {Kwon}\ \emph {et~al.}(2023)\citenamefont {Kwon},
  \citenamefont {Sandhu}, \citenamefont {Shaik}, \citenamefont {Stamm},
  \citenamefont {Sandhu}, \citenamefont {Das}, \citenamefont {Hetherington},
  \citenamefont {Levine},\ and\ \citenamefont {Dantus}}]{kwon2023}%
  \BibitemOpen
  \bibfield  {author} {\bibinfo {author} {\bibfnamefont {S.}~\bibnamefont
  {Kwon}}, \bibinfo {author} {\bibfnamefont {S.}~\bibnamefont {Sandhu}},
  \bibinfo {author} {\bibfnamefont {M.}~\bibnamefont {Shaik}}, \bibinfo
  {author} {\bibfnamefont {J.}~\bibnamefont {Stamm}}, \bibinfo {author}
  {\bibfnamefont {J.}~\bibnamefont {Sandhu}}, \bibinfo {author} {\bibfnamefont
  {R.}~\bibnamefont {Das}}, \bibinfo {author} {\bibfnamefont {C.~V.}\
  \bibnamefont {Hetherington}}, \bibinfo {author} {\bibfnamefont {B.~G.}\
  \bibnamefont {Levine}},\ and\ \bibinfo {author} {\bibfnamefont
  {M.}~\bibnamefont {Dantus}},\ }\bibfield  {title} {\bibinfo {title} {What is
  the mechanism of {H$_{3}^{+}$} formation from cyclopropane?},\ }\href
  {https://doi.org/10.1021/acs.jpca.3c05442} {\bibfield  {journal} {\bibinfo
  {journal} {J. Phys. Chem. A}\ }\textbf {\bibinfo {volume} {127}},\ \bibinfo
  {pages} {8633} (\bibinfo {year} {2023})}\BibitemShut {NoStop}%
\bibitem [{\citenamefont {Wang}\ and\ \citenamefont
  {Yarkony}(2018)}]{wang2018}%
  \BibitemOpen
  \bibfield  {author} {\bibinfo {author} {\bibfnamefont {Y.}~\bibnamefont
  {Wang}}\ and\ \bibinfo {author} {\bibfnamefont {D.~R.}\ \bibnamefont
  {Yarkony}},\ }\bibfield  {title} {\bibinfo {title} {Determining whether
  diabolical singularities limit the accuracy of molecular property based
  diabatic representations: The 1, 2$^{1}${A} states of methylamine},\ }\href
  {https://doi.org/10.1063/1.5048312} {\bibfield  {journal} {\bibinfo
  {journal} {J. Chem. Phys.}\ }\textbf {\bibinfo {volume} {149}},\ \bibinfo
  {pages} {154108} (\bibinfo {year} {2018})}\BibitemShut {NoStop}%
\end{thebibliography}%

\end{document}